\documentclass[aps,pre,twocolumn,eqsecnum,superscriptaddress]{revtex4-2}

\usepackage{color}
\usepackage{hyperref}
\usepackage{amsmath}
\usepackage{graphicx}
\usepackage{color}
\usepackage{wrapfig}
\usepackage[export]{adjustbox}
\begin{document}

\title{The Spanning Tree Model for the Assembly Kinetics of RNA Viruses.}

\author{Inbal Mizrahi}
\affiliation{Department of Physics and Astronomy, University of California, Los Angeles, CA 90095}
\author{Robijn Bruinsma}
\affiliation{Department of Physics and Astronomy, University of California, Los Angeles, CA 90095}
\affiliation{Department of Chemistry and Biochemistry, University of California, Los Angeles, CA 90095}
\author{Joseph Rudnick}
\affiliation{Department of Physics and Astronomy, University of California, Los Angeles, CA 90095}

\date{\today}

\begin{abstract}
We present a simple kinetic model for the assembly of small single-stranded RNA viruses that can be used to carry out analytical packaging contests between different types of RNA molecules. The RNA selection mechanism is purely kinetic and based on small differences between the assembly energy profiles. RNA molecules that win these packaging contests are characterized by having a minimum \textit{Maximum Ladder Distance} and a maximum \textit{Wrapping Number}. The former is a topological invariant that measures the ``branchiness" of the genome molecule while the latter measures the ability of the genome molecule to maximally associate with the capsid proteins. The model can also be used study the applicability of the theory of nucleation and growth to viral assembly, which breaks down with increasing strength of the RNA-protein interaction. 
\end{abstract}

\maketitle

\section{Introduction} \label{sec:intro}

Most viruses have a rod-like or sphere-like shape. Crick and Watson \cite{Crick1956} noted that the volume enclosed by the protein shell (or ``capsid") of a sphere-like virus determines the length of close-packed viral genome molecules. In a similar way, the length of a rod-like capsid scales with the length of the enclosed genome. They exploited the geometrical connection between capsid and genome to relate the size of a virus to the number of encoded genes, and predicted that spherical viruses should have icosahedral symmetry. This approach works well for many smaller single-stranded (ss) RNA viruses (such as the polio and common cold viruses) as well as for some larger double-stranded DNA viruses (such as the herpes and bacteriophage viruses) \footnote{For double-stranded DNA viruses the genome molecules either are inserted into the interior pre-fabricated spherical capsids by a molecular motor or the genome is pre-condensed prior to assembly.}. The genome molecules of the very large retroviruses, like HIV-1, are not close-packed but the dimensions of the capsid still scale with that of the enclosed genome because the genome molecules adhere to the inner surface area of the capsid so the surface area is proportional to the total length of the genome molecules \cite{Ganser-Pornillos2008}. 

Are there other relations between the geometry of that capsid and that of the genome molecules? The \textit{co-assembly} of small single-stranded RNA viruses involves both the capsid proteins and the viral RNA molecules. Such viruses often will self-assemble spontaneously in solutions that contain higher concentrations of viral capsid proteins and RNA molecules \cite{Fraenkel-Conrat1955, Butler1978}. The process is passive and driven by free energy minimization. Early work by Aaron Klug \cite{Klug1999} indicated that the RNA genome molecules act as \textit{templates} that direct the viral assembly process. He proposed a physical model in which repulsive electrostatic interactions between capsid proteins are just strong enough to prevent spontaneous capsid assembly driven by hydrophobic attractive interactions between capsid proteins but if viral RNA molecules are present as well then the negative charges of RNA molecules can neutralize some of the positive capsid protein charges and thereby tilt the free energy balance towards assembly \footnote{For a quantitative treatment of this model, see ref. \cite{kegel2006}}. 

The genome molecules of ss RNA viruses genome molecules have a tree-like ``secondary structure" produced by Watson-Crick base-pairing between complementary RNA nucleotides of the primary sequence of RNA nucleotides \cite{mathews1999}. The co-assembly picture suggests that there could be connections between the geometric and topological features of the viral RNA secondary structure directing assembly and the capsid geometry. The redundancy of the genetic code indeed allows for the possibility of ``silent'' (or synonymous) mutations that can alter the secondary structure of an ss RNA molecule without altering the structure of the proteins encoded in the nucleotide sequence \cite{tubiana}. A systematic comparison between the secondary structures of viral RNA molecules and randomized versions of the same molecules indeed revealed that they are significantly more branched and compact than generic RNA molecules \cite{yoffe2008}. 

The assembly of \textit{empty} viral capsids is similar to that of micelles and other amphipathic molecular systems that self-assemble spontaneously \cite{Safran1994}. Empty capsid assembly may be viewed as a chemical reaction where a certain number of capsid proteins react to form the capsid. Application of the Law of Mass Action (LMA) for reactions in chemical equilibrium produces a relation between the concentrations of assembled capsids and of free capsid proteins with the total protein concentration. This relationship has been confirmed experimentally \cite{Ceres2002} (but see ref.\cite{Morozov2009}). An important feature of the LMA is the presence of a \textit{critical aggregation concentration} (or CAC). For the case of empty capsid assembly, this means that capsids only can form if the total capsid protein concentration exceeds this CAC. Another important feature is that assembled capsids should spontaneously disassemble if the concentration of capsid proteins in the surrounding solution is reduced to zero. This does not occur in actuality---at least not on laboratory time scales---presumably because of the presence of an activation energy barrier inhibiting disassembly that is large compared to the thermal energy $k_BT$. 

Kinetic studies of empty capsid assembly \cite{Prevelige1993, Casini2004, medrano} report that there is a delay or lag time before capsid assembly can take place in a solution of capsid proteins that has been primed for assembly. Assembly starts with the formation of a ``nucleation complex" composed of a small number of capsid proteins. The nucleation complex then extends or elongates by absorbing more proteins until the capsid closes up. The theory of nucleation and growth \cite{Zandi2006} has become an important tool for the interpretation of kinetic studies, with the nucleation complex corresponding to the critical nucleus. The small size of the critical nucleus means that in-vitro assembly assays are taking place under conditions of a significant level of \textit{supersaturation} and, based on kinetic considerations, the same is probably true under in-vivo conditions. A study of the assembly kinetics of empty capsids of the Hepatitis B virus indicates that the assembly of such capsids is governed by distinct \textit{assembly pathways} \cite{asor2019}, much like distinct assembly pathways govern the folding of proteins \cite{bryngelson}. While all-atoms simulations of capsids are possible \cite{Freddolino2006}, coarse-grained models have been found to be useful to interpret equilibrium and kinetic properties of empty capsid assembly\cite{Zlotnick1994, Bruinsma2003, Lidmar2003, Zandi2004, Zlotnick2007, Rapaport2008b, Arkhipov2006, Morozov2009, Mannige2009, Kaplan2014a}.

The co-assembly of capsid proteins with ss RNA molecules involves additional thermodynamic parameters. One of these is the \textit{mixing ratio} of the concentration of RNA molecules to that of the capsid proteins. An in-vitro study \cite{Comas-Garcia} of the co-assembly of CCMV (Cowpea Chlorotic Mottle Virus) with viral RNA molecules reported that when the RNA-to-protein mixing ratio is low then virus-like particles co-exist with excess proteins. By analogy with formation reactions of a binary compound, one would expect that in the case of large mixing ratios there is co-existence of virus-like particles with excess RNA molecules. The border-line mixing ratio separating these two regimes should correspond to the RNA-to-protein ratio of an assembled virus-like particle, which can be thought of as a \textit{stoichiometric ratio}. In actuality, when the mixing ratio exceeded a certain threshold then a distribution of \textit{disordered} RNA-protein aggregates was observed \cite{Comas-Garcia}. Moreover, the border-line mixing ratio was not the stoichiometric ratio. A second thermodynamic parameter found to be important for co-assembly is the ratio of the RNA to protein affinity with the protein to protein affinity. This affinity ratio can be altered experimentally by changing the acidity and salinity of the solution \cite{Garmann}. It was found that for increased values of this ratio, virus-like particles were replaced by disordered RNA-protein aggregates.

Information about co-assembly also has been gleaned from structural studies. Until recently, reconstruction of packaged genome molecules involved ``icosahedral averaging", which resulted in RNA structures with imposed icosahedral symmetry \cite{Baker}. Such studies showed that the interior surface of the icosahedral capsids of certain viruses (e.g., the nodaviruses \cite{Tihova2004}) is decorated by paired RNA strands lining the edges of the ``capsomers" (i.e., pentameric or hexameric groupings of capsid proteins)).  Recent progress in cryo-electron tomography has made it possible to image individual ss RNA genome molecules packaged inside spherical capsids without icosahedral averaging (``asymmetric reconstruction"\cite{koning, beren}). One example is the MS2 virus, which is an ss RNA, quasi-icosahedral bacteriophage virus. It was found that a subsection of the RNA genome reproducibly associated with a compact cluster of capsid proteins \cite{Dykeman2011}. This result was interpreted as evidence for a well-defined assembly pathway where the energetically "uphill" part of nucleation-and-growth scenario produces a compact nucleation complex held together by particular sections of the viral RNA molecule with enhanced affinity for the capsid proteins (such sections are known as \textit{packaging signals} \cite{Patel2015}). In this picture, the ``downhill" part of the assembly process involves more generic electrostatic RNA-protein attractions. According to this model, the nucleation complex plays the important role of selecting the viral RNA molecules from quite similar host messenger RNA molecules.

In this paper we propose a statistical mechanical model that can be used to explore the question of the influence of geometric and topological features of tree-shaped molecules on the promotion of packaging by capsid proteins and how these features are connected to kinetically favored assembly pathways. It allows testing of the theory of nucleation and growth. Other questions that can be addressed are why the ``chemical reaction picture" works well for empty capsid assembly but not so well for co-assembly, whether packaging selectivity is consistent with a high level of supersaturation. 

The proposed model, the\textit{``spanning tree model"} model, extends a simple model for the assembly of empty dodecahedral capsids from pentamers in solution, due to Zlotnick \cite{Zlotnick1994, Endres2002}, by allowing it to package branched genome molecules that decorate the edges of the dodecahedral capsid. An important advantage of the Zlotnick Model is that its assembly kinetics has been studied \cite{Zlotnick2007, Morozov2009} while the packaging kinetics of linear genome molecules has been numerically simulated \cite{Perlmutter2014}. Two different packaging scenarios were encountered. The first scenario is similar to that of the assembly of empty capsids with assembly intermediates in the form of compact pentamer clusters that grow in a pentamer-by-pentamer fashion. In the second ``en masse" scenario, the first assembly step is the formation of a disordered pentamer/genome condensate formed which then undergoes an ordering transition that can be described by the Landau theory of symmetry-breaking on a spherical surface (see \cite{rudnick2019} and references therein).

\section{The Spanning Tree Model.}
We start with a brief review of the Zlotnick Model \cite{Zlotnick1994}.
\subsection{Zlotnick Model}
\begin{figure}[htbp]
\begin{center}
\includegraphics[width=2.5in]{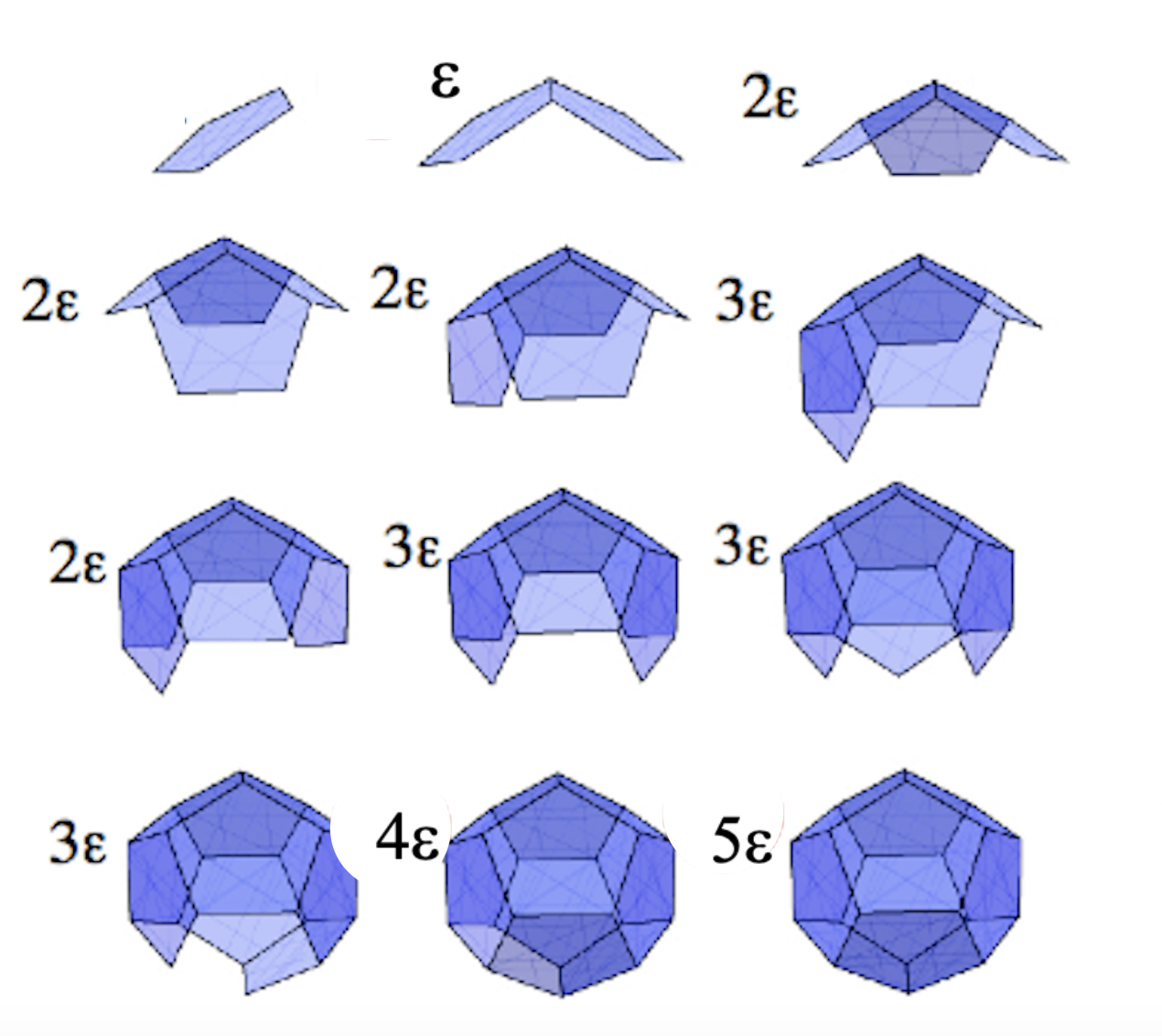}
\caption{Zlotnick Model. The figure shows a minimum-energy pathway for the assembly of a dodecahedral shell composed of twelve pentamers with adhesive edges. The edge-edge binding energy is $\epsilon$. The change in energy per added pentamer is indicated.}
\label{fig:Zlotnick Model}
\end{center}
\end{figure}

In the Zlotnick Model, a capsid is modeled as a dodecahedral shell composed of twelve pentamers. Assembly takes place in a reservoir of pentamers and is driven by an attractive edge-edge interaction between pentamers. A \textit{minimum-energy assembly pathway} is defined as a pentamer-by-pentamer addition sequence where each added pentamer is placed in a location on a partially assembled dodecahedral shell that minimizes the total energy. An example of a minimum-energy assembly pathway is shown in Fig. \ref{fig:Zlotnick Model}.   

The energy $E(n)$ of a cluster of $n$ pentamers is defined as the number of shared pentamer edges times the binding energy $\epsilon$ per edge minus a constant $\mu_0$ times the number of pentamers. Here, $\mu_0$ is the chemical potential of a pentamer in solution at a certain reference pentamer concentration $\bar{c}$ \footnote{The assembly of the capsid is assumed here to take place on a specific location. The quantity $\mu_0$ reflects the entropic free energy cost of removing a pentamer from the solution to this location plus that of any conformational change that is required for the pentamer prior to joining a partial capsid.} in units of $k_BT$. If the concentration of free pentamers $c_f$ differs from $\bar{c}$ then $\mu_0$ must be replaced by $\mu=\mu_0 + \ln c_f/\bar{c}$ but in this section $c_f$ equals $\bar{c}$. From here on, concentrations are expressed in terms of $\bar{c}$ and thus dimensionless. Since a dodecahedron has thirty edges, the assembly energy of a complete capsid equals $30\epsilon-12 \mu_0$. If the chemical potential equals $(5/2)\epsilon$ then the assembly energy of a pentamer that is part of a capsid is the same as that of a free pentamer in solution (i.e., zero). We will refer to $\mu^*=(5/2)\epsilon$ as the chemical potential for assembly equilibrium. Below, we will use energy units in which $\epsilon=-1$ so with $\mu^*$ equal to -(5/2). An important feature of the Zlotnick model is that the energy gain for the initiation of assembly is relatively low, namely one factor of $\epsilon$, while the cost of removing a single pentamer from an assembled capsid is relatively high: it requires breaking five bonds with a total energy cost of minus $5\epsilon$. This allows for metastability of assembled capsids in solutions with low pentamer concentrations.

\begin{figure}[htbp]
\begin{center}
\includegraphics[width=3.5in]{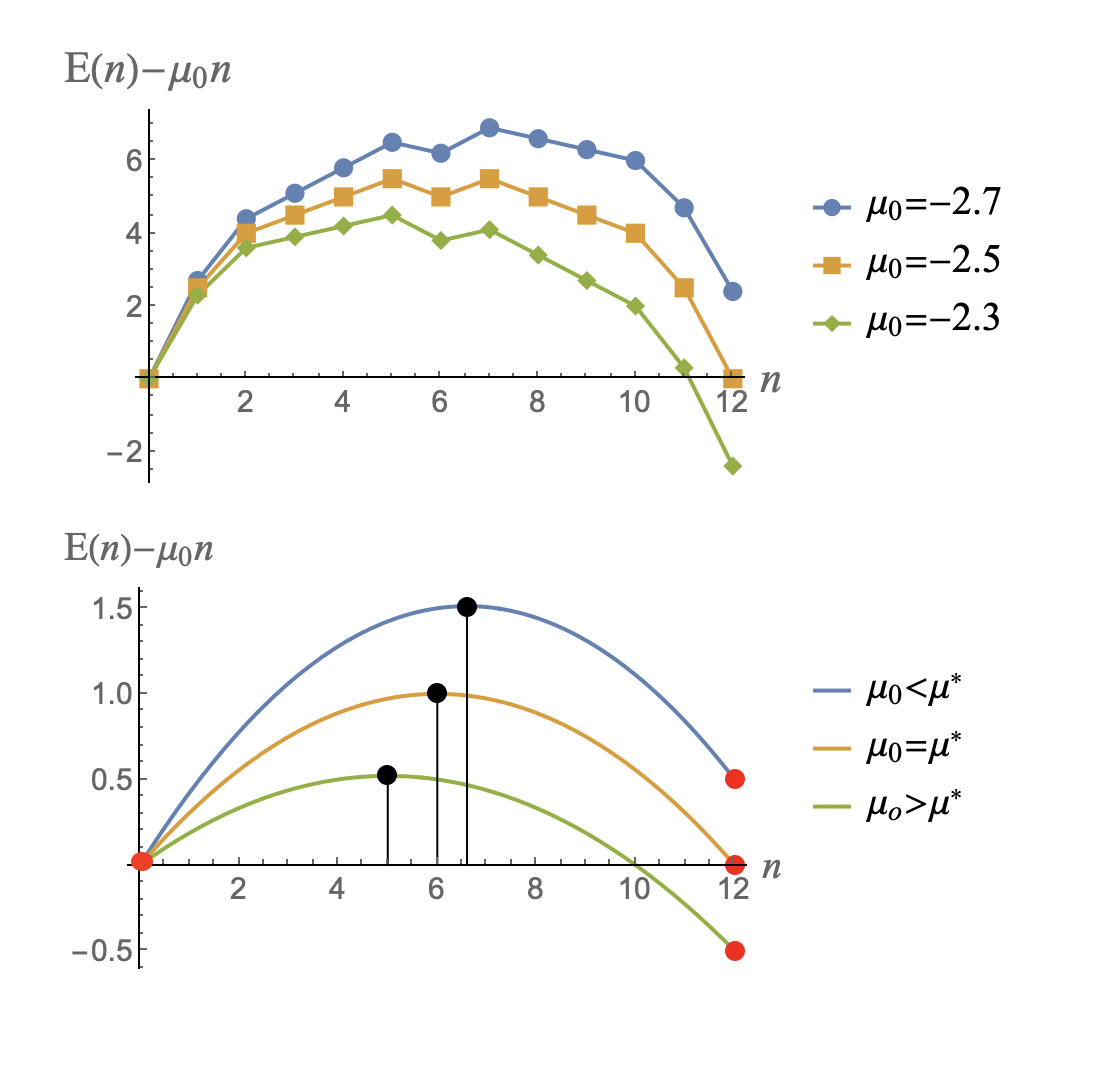}
\caption{Top: Assembly energy profile $E(n)$ for the minimum assembly energy pathway of the Zlotnick Model shown in Fig. \ref{fig:Zlotnick Model}. Blue dots: $\mu_0$ is slightly below the chemical potential $-(5/2)$ for assembly equilibrium. Orange squares: $\mu_0$ is equal to $-(5/2)$. Green diamonds: $\mu_0$ is slightly above $-(5/2)$. The edge-to-edge binding energy is equal to minus one. 
Bottom: interpretation of the energy profile in terms of nucleation-and-growth theory. Solid red dots: energy minima corresponding to the assembled capsid. Solid black dots: energy maxima. For $\mu_0<\mu^*$, the absolute energy minimum is at $n=0$ while for $\mu_0>\mu^*$ the absolute minimum is at $n=12$, the assembled capsid. The height of the maxima corresponds to the activation energy barrier. The location $n^*$ of the maximum is around $n=6$ in the assembly equilibrium state and shifts to smaller values as $\mu_0$ increases. }
\label{fig.2}
\end{center}
\end{figure}

Figure \ref{fig.2} (top) shows the minimum-energy assembly pathway corresponding to Fig. \ref{fig:Zlotnick Model} for three different values of $\mu_0$ close to $\mu^*$ and for $\epsilon=-1$.
There is a large number of such minimum energy assembly pathways for the Zlotnick model (of the order of $10^5$) with the energy profile of Fig. \ref{fig.2} (top). Figure \ref{fig.2}(bottom) shows how these energy profiles can be interpreted in terms of the theory of nucleation and growth \cite{Zandi2006}. The height of the energy maximum plays the role of the energy activation barrier. The size of the critical nucleus is the value $n^*$ for which $E(n^*)$ has a maximum. If $\mu_0\simeq\mu^*$, then $n^*$ is around 6 while $n^*$ decreases with increasing levels of pentamer supersaturation. Note that the Zlotnick Model effectively incorporates the surface or line tension that plays a key role in the theory of nucleation and growth. 

For the Zlotnick Model, as well as other simple models of capsid assembly, the critical nucleus under assembly equilibrium conditions ($\mu_0=\mu^*$) is a \textit{half-formed capsid}. Experimentally measured values for the interaction energies between capsid proteins are in the range of a few $k_BT$. If the critical nuclei really were half-formed capsids then the activation energy barrier for the assembly of actual capsids would be in the range of hundreds of $k_BT$ under conditions of assembly equilibrium, which would mean prohibitively slow kinetics. Actual measured values of the size of the nucleation complex are much smaller \cite{Casini2004}. As already noted, this indicates that capsid assembly takes place under conditions of a high degree of supersaturation. 

\subsection{Spanning Trees}
\begin{figure}[htbp]
\begin{center}
\includegraphics[width=3in]{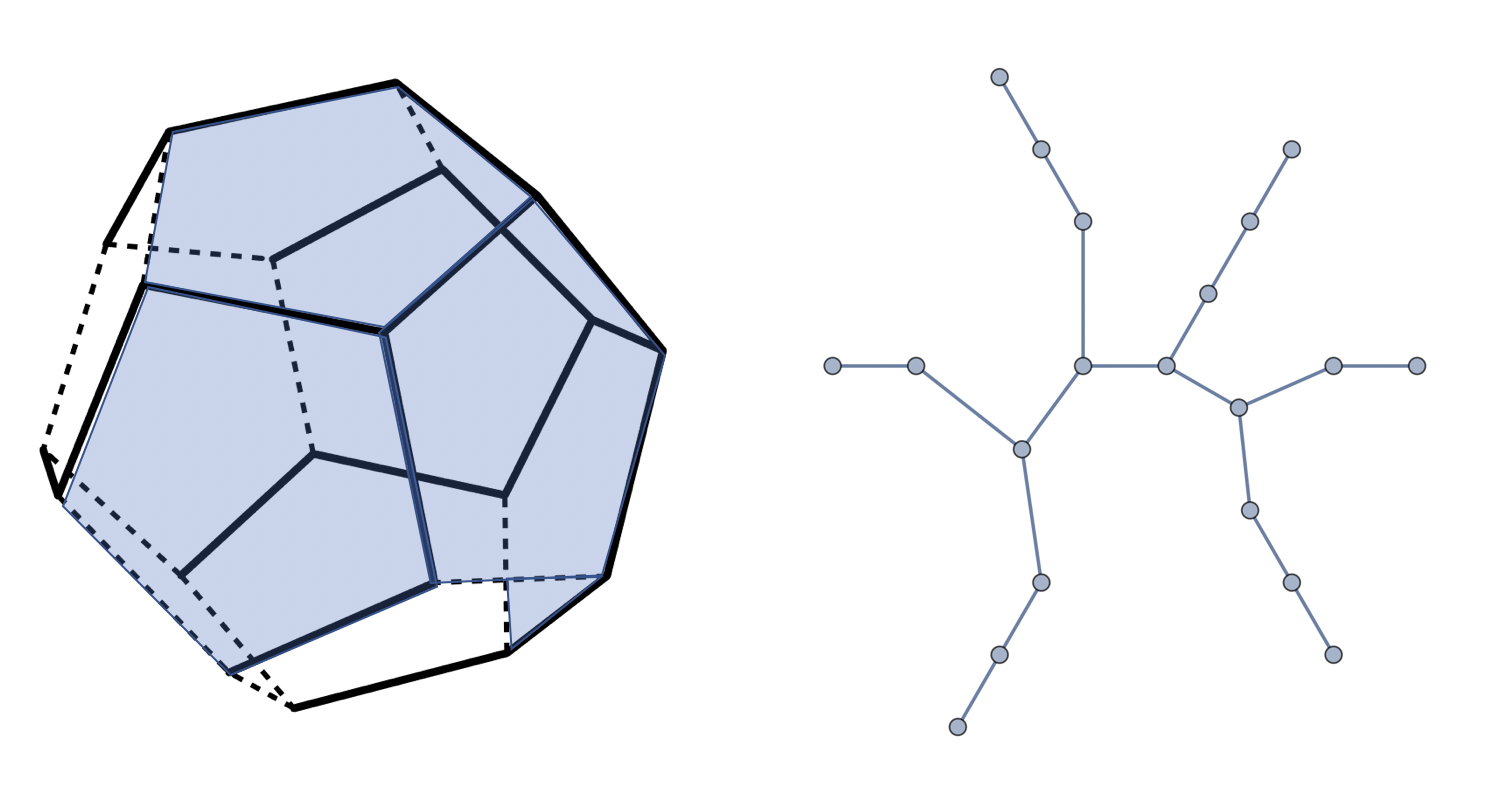}
\includegraphics[width=2in]{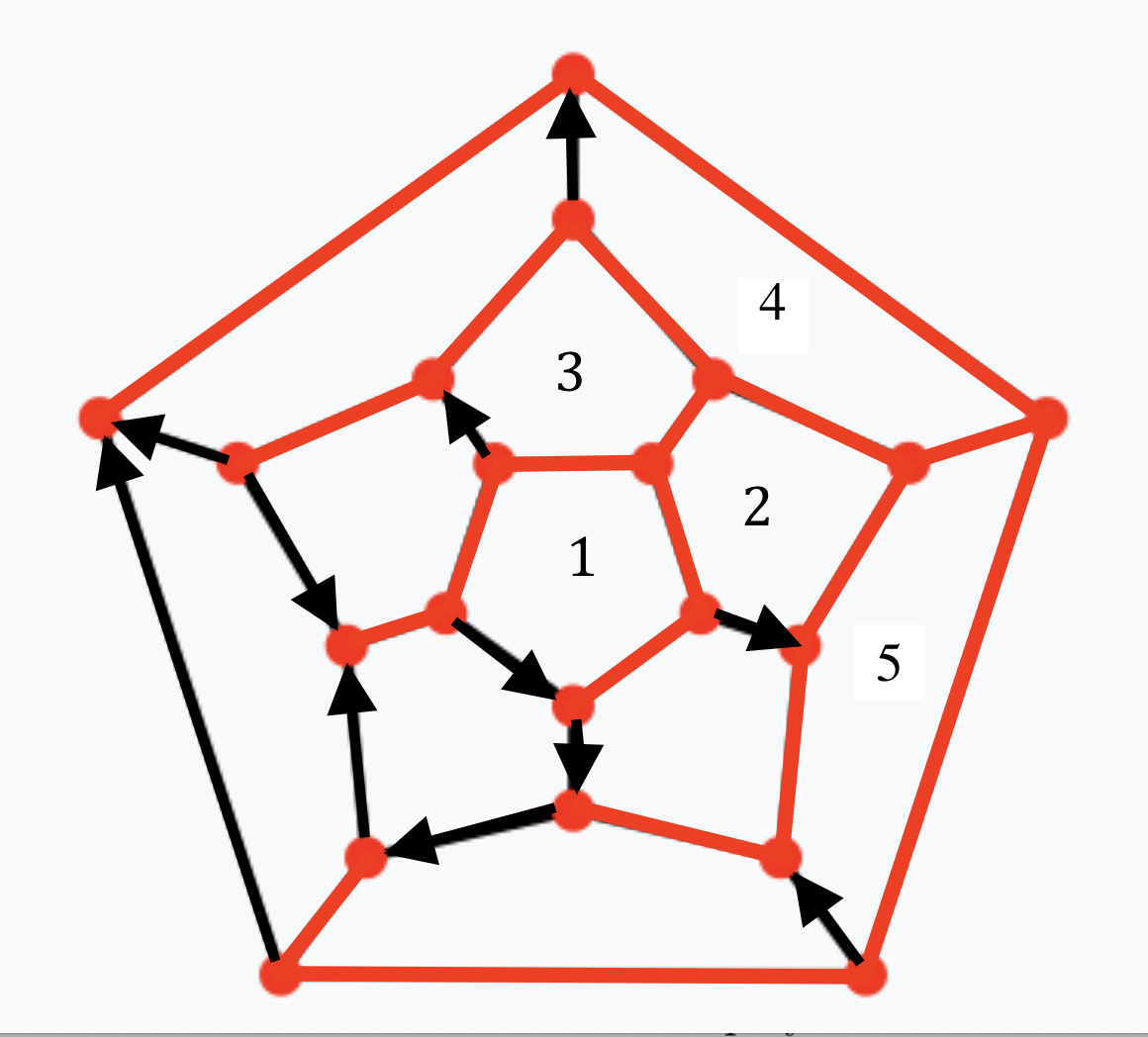}
\caption{Top left: Spanning tree connecting the vertices of a dodecahedron (solid lines). The dashed lines indicate edges of the dodecahedron that are not part of the spanning tree of specific links. Six pentamers can be placed on the dodecahedron with each one wrapped by the spanning tree with four links per pentamer (numbered). Top right: Planar graph of the same tree. Bottom: The spanning tree projected on a planar Schlegel graph of the dodecahedron (red). The black arrows represent side branches that are added to the spanning tree so that all edges of the dodecahedron are covered by a link of the tree.}
\label{fig:dodtrees}
\end{center}
\end{figure}

The second part of the definition of the model is the specification of ``toy" ss RNA genome molecules. These are represented by \textit{tree graphs}, i.e., collections of nodes connected by links such that there is one and only one path of links connecting any pair of nodes \cite{Bollobas} (see Fig. \ref{fig:dodtrees} top right). The construction of a genomic tree graphs starts with a \textit{spanning tree graph} of the dodecahedron. A spanning tree graph of a polyhedron is defined as a tree graph whose nodes are located on the vertices of the polyhedron with just enough links to connect the nodes together in a tree structure without circuits \cite{Graham}. Each member of the set of spanning trees of the dodecahedron has the same number of vertices (twenty) and the same number of links (nineteen). Figure \ref{fig:dodtrees} shows three ways to represent the same spanning tree graph. 
The top left figure shows a three dimensional representation with the solid black lines indicating the spanning tree. Note that the spanning tree covers only 19 of the 30 edges of the dodecahedron. The top right figure is a two dimensional representation of a tree graph with the dodecahedron removed. In the bottom picture, the dodecahedron is represented in the form of a planar Schlegel diagram \cite{loeb} that is decorated by the spanning tree graph (solid red lines). The final step of the construction of a genome molecule is the extension of a spanning tree by adding side-branches so the tree covers the remaining eleven edges. These are represented by black arrows in Fig. \ref{fig:dodtrees}. The completed genomic tree molecule covers all edges of the dodecahedron. The red spanning tree links will represent the packaging signals that have a specific binding affinity to pentamer edges that is enhanced with respect to generic electrostatics while the black arrows represent side branches with only non-specific generic affinity for pentamer edges.

\subsection{Classification Indices for Spanning Tree Molecules.}
 
\begin{figure}[htbp]
\begin{center}
\includegraphics[width=3in]{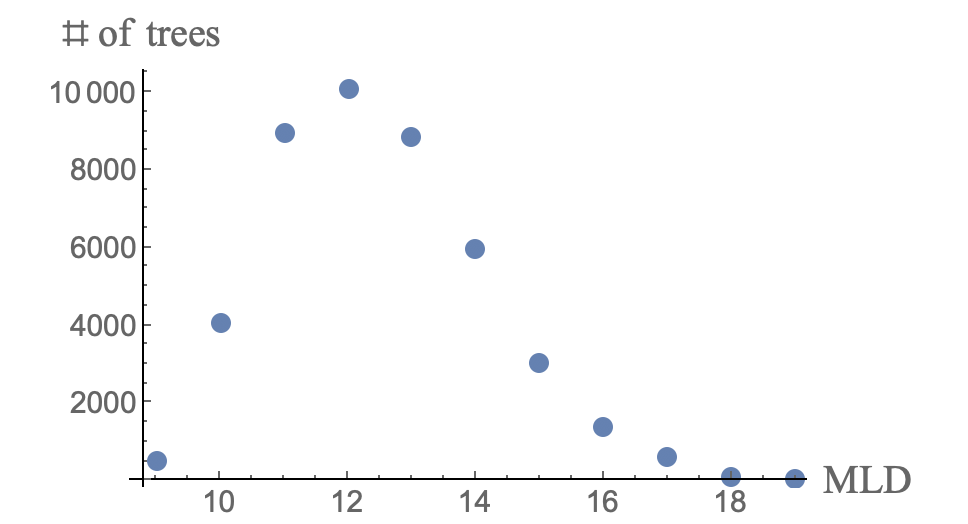}
\caption{The number of spanning trees on the dodecahedron as function of the maximum ladder distance (MLD). Two spanning trees of the dodecahedron that are related by a symmetry operation of the dodecahedron are treated as the same. }
\label{fig:dodlogplot}
\end{center}
\end{figure}

The number of unique spanning trees \footnote{The set of unique spanning trees consists of all spanning trees that, when depicted as in the top left of Fig. \ref{fig:dodtrees}, cannot be mapped into each other by rotations and/or reflections that leave the dodecahedron invariant.  } of the dodecahedron is on the order of $10^5$ and we need to classify them. A ``global" characteristic that has been applied to classify RNA secondary structures is the Maximum Ladder Distance (or MLD)  \cite{yoffe2008, fang}. This is the maximum number of paired RNA nucleotides separating any two nucleotides. The MLD of a secondary structure is a global measure of its size \cite{yoffe2008, fang}. Specifically, in the absence of interaction between nodes, the solution radius of gyration of the molecule scales with the MLD as a power law. The MLD is a topological invariant that does not change if the molecule is folded up in different ways. A systematic comparison between the genomic RNA molecules of RNA viruses revealed that they have significantly lower MLDs than randomized versions of the same molecules \cite{yoffe2008, fang}. The analog of the MLD for the toy genome molecules is the maximum number of links separating any pair of nodes. In graph theory, the ladder distance between two nodes of a tree graph is called ``the" distance while the MLD is known as the ``diameter" of the tree graph \cite{Bollobas}. The MLD of the tree molecule shown in Fig. \ref{fig:dodtrees} is nine. It can be demonstrated that the smallest possible MLD for a spanning tree of the dodecahedron is nine (as shown in Appendix \ref{app:A}), while the largest possible MLD of a spanning tree is nineteen. The former resemble a Bethe lattice while the latter is a Hamiltonian Path. A walk that visits all vertices of a polyhedron is said to trace a \textit{Hamiltonian Path} \cite{Rudnick2005, Dykeman2013b}. Figure \ref {fig:dodlogplot} is a plot of the number N of spanning trees of the dodecahedron as a function of the MLD.
The plot has a pronounced maximum around MLD equal to 12. By comparison, the configurational entropy of an annealed branched polymer composed of 19 monomers that is not constrained to be a spanning tree depends on the MLD as $19-MLD^2/19$ \cite{gutin} and thus has a maximum at the smallest possible MLD. Demanding that a tree molecule with a certain number of links is also a spanning tree of a dodecahedron constrains significantly the branching statistics.

A second characteristic, complementary to the MLD, is the \textit{wrapping number} (or $``N_P"$). The wrapping number of a spanning tree of the dodecahedron is the maximum number of pentamers that can be placed on the dodecahedron such that all pentamers have four edges covered by a link of the spanning tree (four---not five---is the maximum number of \textit{specific} links that can be associated with a pentamer). For the spanning tree shown in Fig. \ref{fig:dodtrees}, $N_P$ equals six. The maximum $N_P$ for a spanning tree of the dodecahedron is eight while the minimum is two. The distribution of wrapping numbers is shown in Fig. \ref{fig:NP}:
\begin{figure}[htbp]
\begin{center}
\includegraphics[width=3in]{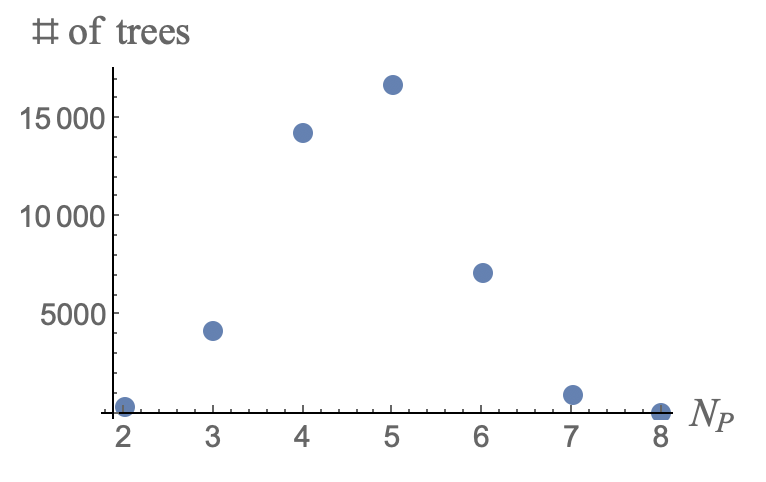}
\caption{The number of spanning trees on the dodecahedron as a function of the wrapping number. Two spanning trees of the dodecahedron that are related by a symmetry operation of the dodecahedron are treated as the same. }
\label{fig:NP}
\end{center}
\end{figure}
The wrapping number distribution has a maximum at $N_P=5$. 

\begin{figure}[htbp]
\begin{center}
\includegraphics[width=2.0in]{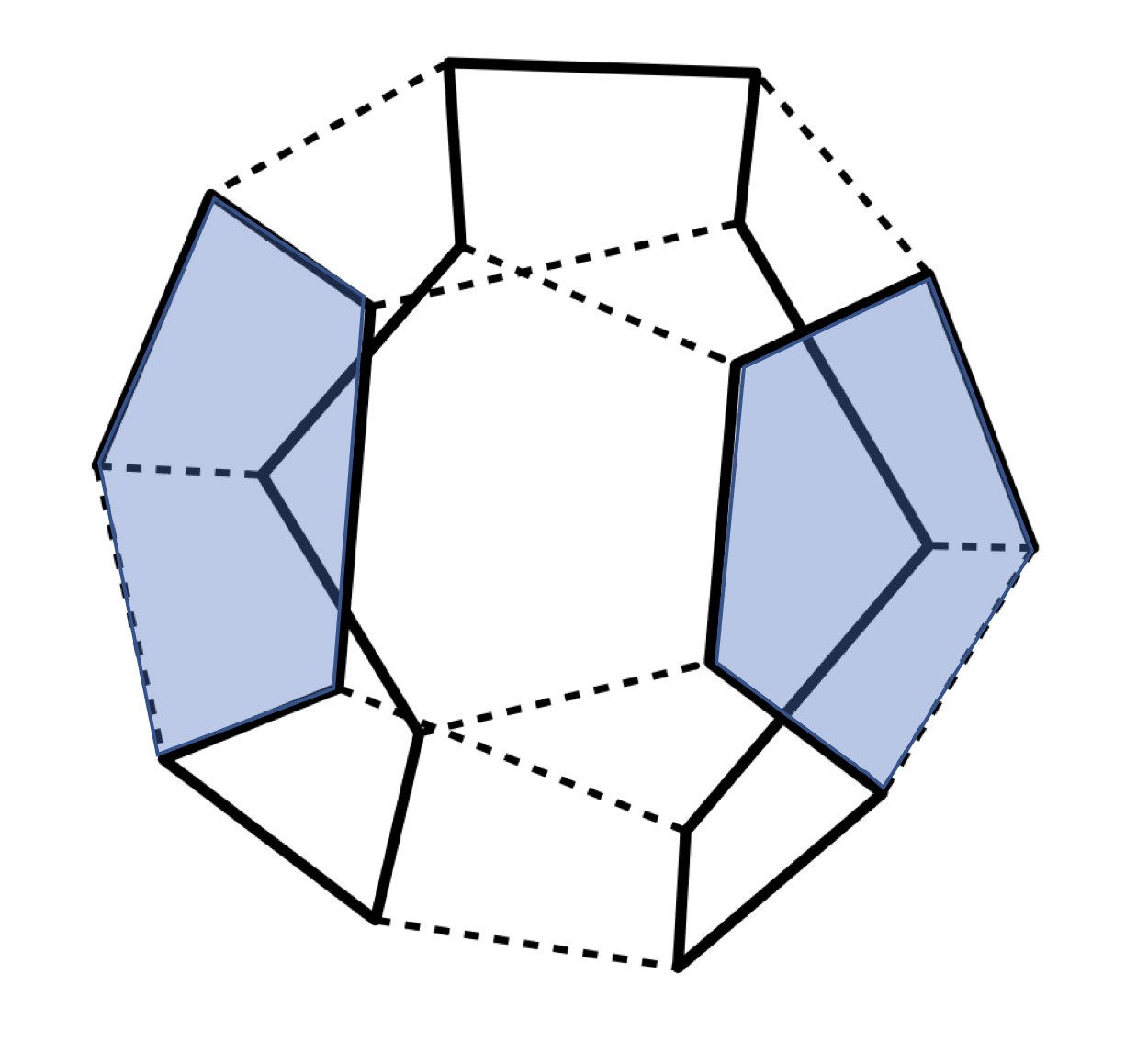}
\caption{A possible Hamiltonian path for a linear genome molecule. Only two of the pentamers are maximally wrapped. They have no shared edge.}
\label{fig:HP}
\end{center}
\end{figure}

\begin{figure}[htbp]
\begin{center}
\includegraphics[width=3.5in]{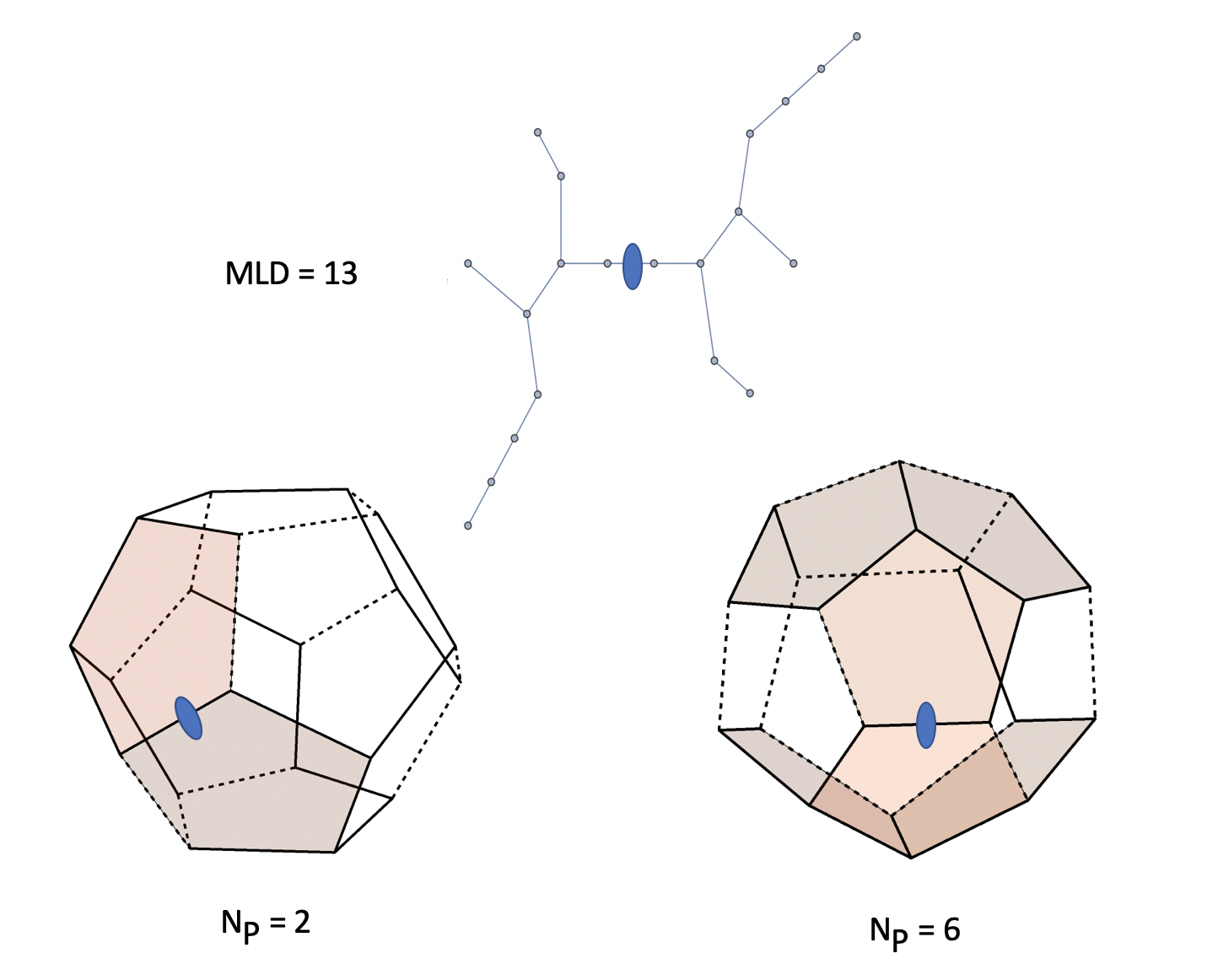}
\caption{Top: An MLD=13 tree molecule with a two-fold symmetry site (marked). Bottom: Two dodecahedral spanning tree configurations of this molecule. Both configurations retain two-fold symmetry (marked). The configuration on the left has a wrapping number $N_P=2$ while the one on the right has a wrapping number $N_P=6$}
\label{fig:Deg}
\end{center}
\end{figure}

Unlike the MLD, the wrapping number of a tree molecule is not an invariant of the tree topology. Instead, it depends on the spatial configuration of a tree molecule distributed over the edges of a dodecahedron. An example is a \textit{linear} spanning tree as shown in Fig. \ref{fig:HP}. 
The Hamiltonian path of Fig. \ref{fig:HP} has a wrapping number of two. After enumerating all possible Hamiltonian paths, one finds that their wrapping number can be two, three, and four. In general, a spanning tree molecule can be placed in different ways on the edges of a dodecahedron and these different configurations may have different wrapping numbers. Figure \ref{fig:Deg} shows two spanning tree configurations with different wrapping numbers that belong to the same tree molecule. 
Tree structures can have multiple wrapping numbers. The wrapping number is a geometrical characteristic of the different ways to distribute nineteen specific links over the edges of a dodecahedron. A network with circuits that would visit all nodes of a dodecahedron would not have a well-defined MLD but it would still have a wrapping number.
 
The wrapping number and MLD are correlated. For example, a spanning tree with a large wrapping number is expected to be have many branches in order to maximize the number of pentamers that can be accommodated. Hence it is expected to have a small MLD (and vice versa). Figure \ref{fig:NP_MLD} is a plot of the range of allowed wrapping numbers for given MLD: 
\begin{figure}[htbp]
\begin{center}
\includegraphics[width=3in]{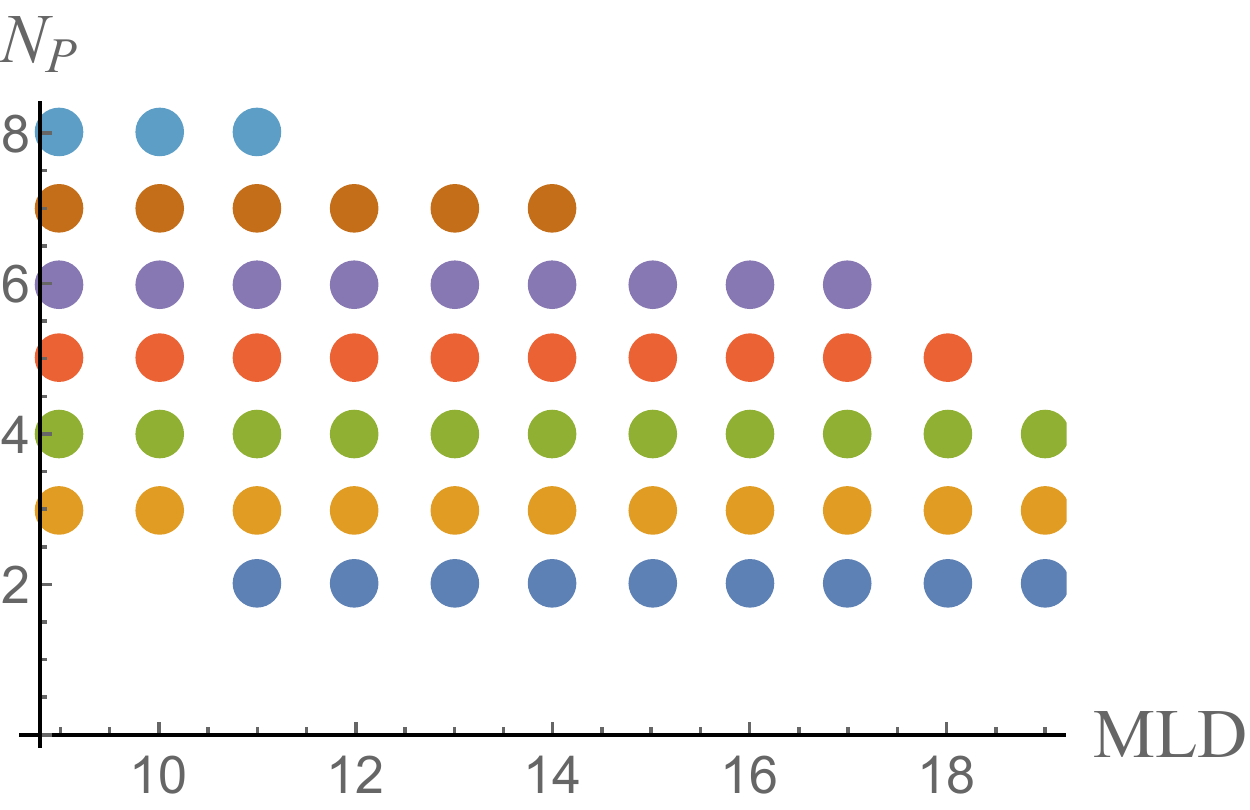}
\caption{Plot of the range of wrapping numbers ($N_P$) of the spanning tree molecules of the dodecahedron as a function of the maximum ladder distance (MLD).}
\label{fig:NP_MLD}
\end{center}
\end{figure}
The largest possible wrapping number is $N_P=8$. In that case, the MLD can only be nine, ten or eleven.  For the smallest possible wrapping number $N_P=2$ the MLD ranges from eleven to nineteen. 

\subsection{Assembly Energy Profiles.}

We now construct the minimum energy assembly energy profiles for the spanning tree model. The starting state is a tree molecule, composed of a spanning tree (the specific links) plus the additional eleven side branches (the non-specific links), that decorates all edges of a mathematical dodecahedron. Physically, the starting state can be viewed as representing a folded or pre-condensed form of the viral ss RNA genome molecule(s) prior to its encapsidation by pentamers \footnote{In actuality, condensation of the RNA genome molecules takes place \textit{during} encapsidation. It is driven by positively charged polypeptide chains associated with the capsid proteins.} Next, place $n$ pentamers on the dodecahedron. The energy $E(n)$ of the assembly is defined to be $E(n) = n_1\epsilon_1 + n_2\epsilon_2+n_3\epsilon_3 + n_4\epsilon_4-k_BT\mu_0 n$. Here, $n_1$ is the number of specific links of the tree that lie along a pentamer edge that is not shared with another pentamer, $n_2$ is the number of specific links that lie along a pentamer edge that is shared with another pentamer, $n_3$ is the number of edges shared between two pentamers that are associated with a non-specific link and $n_4=11$ is the number of non-specific links that lie along a pentamer edge that is not shared with another pentamer. The associated binding energies are given as $\epsilon_i$ with $i =1,2,3,4$. We will use an energy scale in which the binding energy $\epsilon_4$ of a non-specific link is equal to zero.  We also will assume that the interactions between edges and links are \textit{additive}. This means that $\epsilon_2$ is given by $\epsilon_2=\epsilon_3+2\epsilon_1$.  Finally, $\mu_0$ is again the reference pentamer chemical potential. The assembly energy of a completed particle is equal to $19\epsilon_2+11\epsilon_3-12 \mu_0$. \textit{All} spanning trees thus have the same assembly energy. If $\epsilon_1$ is zero so $\epsilon_2=\epsilon_3$ then the assembly energy profiles are the same as that of the Zlotnick Model with $\epsilon=\epsilon_2=\epsilon_3$. 

The energy parameters enter in the physics of assembly discussed in the next sections always in the form of $\beta\epsilon_i$ with $\beta=1/k_BT$. We will use dimensionless energy parameters $\epsilon'_i = \epsilon_i/|\epsilon_3|$. In these units $\epsilon'_3=-1$ and $\beta'=\beta/|\epsilon_3|$. That leaves only $\epsilon'_1$, $\beta'$ and $\mu_0$ as three free dimensionless energy parameters. Physically, $|\epsilon'_1|$ represents the ratio between the affinities of a specific link with a pentamer edge and that of pentamer edges with each other $1/\beta'$ is the dimensionless attractive interaction strength between two pentamers in units of the thermal energy. Below we will drop the primes.

\begin{figure}[htbp]
\begin{center}
\includegraphics[width=3.5in]{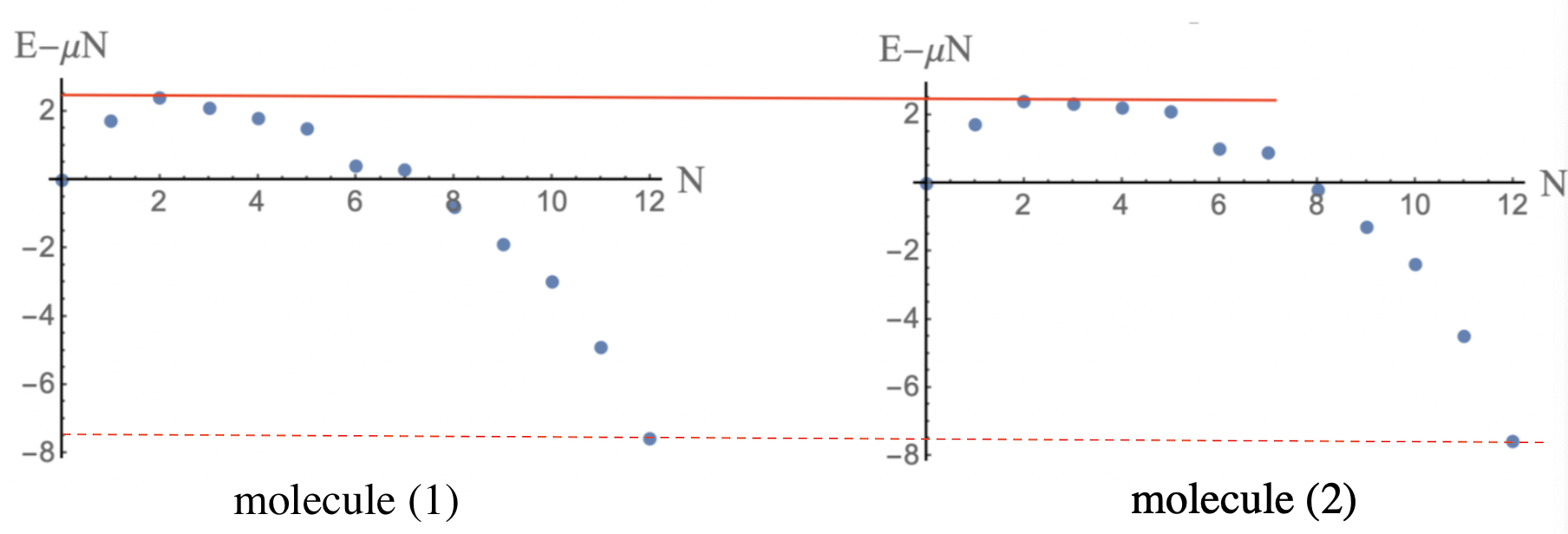}
\caption{Assembly energy profiles for minimum energy assembly pathways for the $MLD=9$, $N_P=5$ spanning tree of Fig. \ref{fig:dodtrees} (left) and the $MLD=19$, $N_P=2$ spanning tree of Fig. \ref{fig:HP}. Energy parameters are $\epsilon_1=-0.2$, $\epsilon_3=-1$, and $\mu_0=-2.5$. The total assembly energy is indicated by a dashed red line and the assembly activation barrier by a solid red line.}
\label{example2}
\end{center}
\end{figure}

Minimum energy assembly pathways can now be constructed in the same way as before. The degeneracy of the energy profile Zlotnick Model is greatly reduced. As an example, Fig. \ref{example2} shows a minimum energy assembly profiles for the $MLD=9$, $N_P=6$ tree of Fig. \ref{fig:dodtrees} (from here on referred to as ``molecule (1)") and for the $MLD=19$, $N_P=2$ tree of Fig. \ref{fig:HP} (from here on referred to as ``molecule (2)"). 
These two energy profiles are consistent with what is expected from nucleation and growth theory for a state of elevated supersaturation. Note how similar they are: the total assembly energies (solid red lines) are of course the same by construction (about 7.5 in our units) but the heights of the assembly energy activation barriers also are the same (about 2.5 in our units). The main difference is that the width of the energy barrier for molecule (2) is somewhat larger than that of molecule (1). Mathematically, the assembly process can be viewed as a random walk over energy landscapes of the form shown in Fig. \ref{example2}, which suggests that molecule (1) will have somewhat faster assembly kinetics.

\begin{figure}[htbp]
\begin{center}
\includegraphics[width=3.5in]{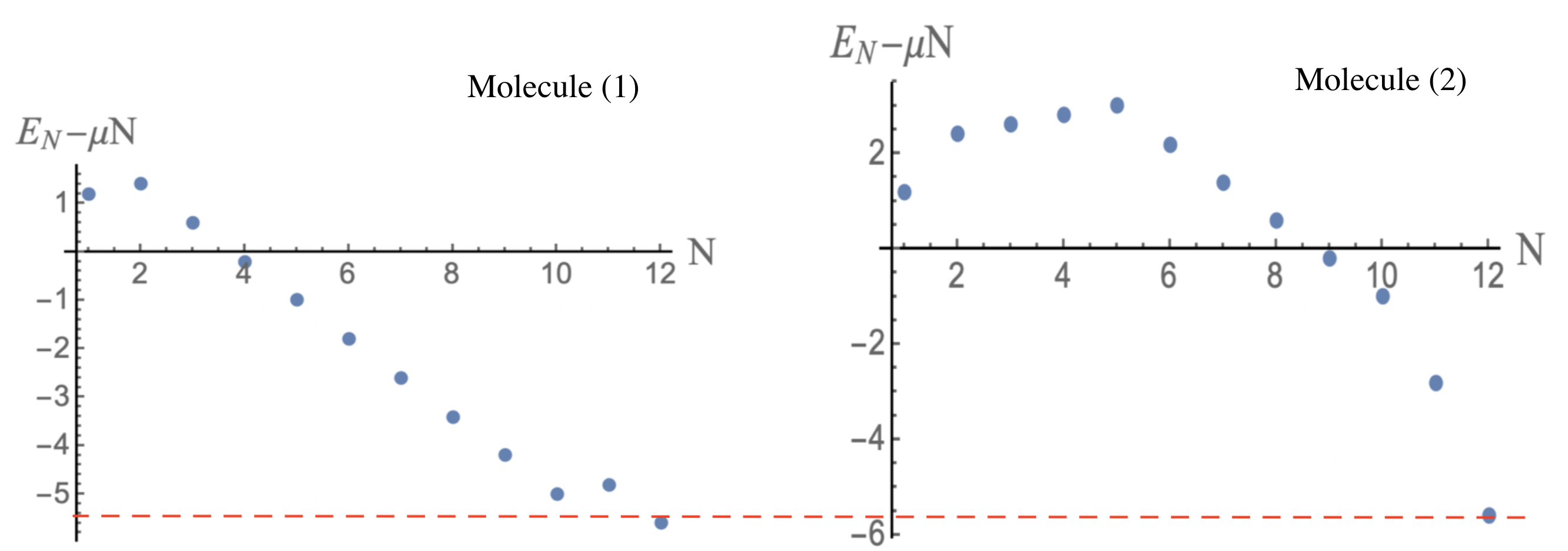}
\caption{Assembly energy profiles for molecule (1) (left) and molecule (2) (right) for energy parameters $\epsilon_1=-1.0$, $\epsilon=-1$, and $\mu_0=-5.2$}
\label{RSN}
\end{center}
\end{figure}

Next, we increased minus $|\epsilon_1|$ from $0.2$ to $1.0$ so the attractive interaction between a tree link and a pentamer edge is equally strong as that between two pentamer edges. We also decreased the reference chemical potential from $-2.5$ to $-5.2$ in order to keep the activation energies roughly in the same range as for $\epsilon=-0.2$. The assembly energy profiles are shown in Fig. \ref{RSN}. 
The total assembly energy is still the same (which is by construction) but the two profiles now have a quite different appearance. The activation energy barriers also are quite different The reason is that the minimum energy placement for the second pentamer for molecule (2) is, for large values of minus $\epsilon_1$ determined by the requirement that it is maximally wrapped. The two maximally wrapped pentamers of molecule (2) do not have a shared edge (see Fig. \ref{fig:HP}). The $n=2$ state of molecule (2) has a higher energy than that of molecule (1) because it lost the attractive energy between the edge shared between the two first pentamers of molecule (1). Examining the assembly energy profiles of molecules with different MLD and $N_P$ we found that, in most cases, genome molecules with the same MLD and $N_P$ have the same, or closely similar, minimum energy assembly profiles and pathways. On the other hand, genome molecules with the same MLD but different $N_P$ can have quite different profiles and pathways. 

Finally, one also can define \textit{disassembly} energy pathways using the same method except that now pentamers are removed successively, each time with the minimum energy disassembly cost. By enumeration we found that, in nearly all cases, the disassembly path simply reverses the assembly path.

\subsection{Mechanical Rigidity.}

Figure \ref{fig:dodtrees} (top left) shows a six pentamer structure placed on the dodecahedron. Each of these pentamers has the maximum of four contacts with a specific (red) link per pentamer. For \textit{any} (negative) value of $\epsilon_1$, the energy $E(6)$ of this state has the lowest value \textit{any} six-pentamer structure could have when placed on the dodecahedron for any spanning tree. In other words, for an $n^*=6$ nucleation complex (with $\mu_0 = \mu^*$) the activation energy barrier could not be any lower. One thus might expect the tree structure of Fig. \ref{fig:dodtrees} to be a particularly strong contender in a packaging competition experiment. This structure has however another interesting feature. If one were to (i) relax the constraints that keep the links of the genome molecule associated with the edges of a mathematical dodecahedron, (ii) allow tree links to rotate freely around the nodes of the tree and (iii) allow pentamers to swivel around shared edges then the pentamers \textit{still} could not move with respect to each other without breaking pentamer-pentamer bonds. Only rigid-body translations and rotations would be possible. We will say that pentamer assemblies with this property are mechanically \textit{rigid}. The pentamer structures of the minimum energy assembly pathway of the Zlotnick Model (see Fig.1) are all rigid past $n=2$.

\begin{figure}[htbp]
\begin{center}
\includegraphics[width=2.0in]{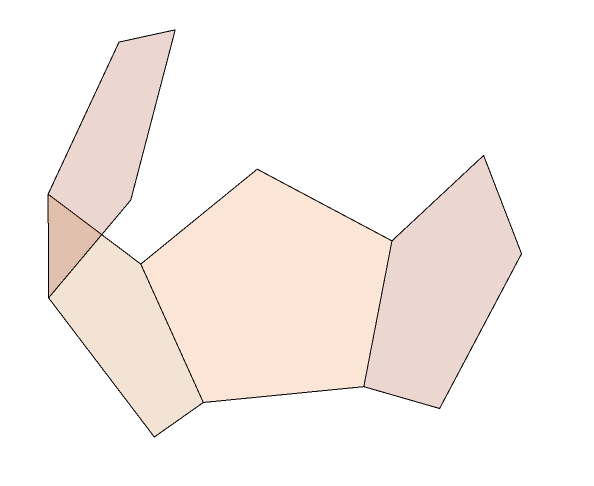}
\caption{Intermediate $n=4$ assembly state of molecule (2) for large values of $|\epsilon_1|$. The structure is mechanically flaccid.}
\label{fig:F}
\end{center}
\end{figure}

 As a counterexample, the linear genome molecule with 19 edges of Fig. \ref{fig:HP} with wrapping number two can accommodate only two pentamers with a maximum of four favorable contacts. Placing a third and fourth pentamer on the dodecahedron along a minimum energy assembly path generates the linear four-pentamer structure shown in Fig. \ref{fig:F}. 
If this structure were released from the dodecahedron and allowed to freely fluctuate then the four pentamers could freely swivel along shared edges. We will call such an unstable assembly mechanically \textit{flaccid}. Assembly intermediates that are flaccid are expected to be characterized by strong conformational thermal fluctuations. Whether or not an assembly intermediates are flaccid is controlled by the parameter $\epsilon_1$. We find that in the limit that $\epsilon_1$ is small compared to the other energy parameters, all assembly intermediates are rigid and basically reproduce the scenario of the Zlotnick Model. The assembly intermediates of some---though not all---pathways become flaccid when $|\epsilon_1|$ is increased past a critical value of the order of one. For molecule (2) the transition from rigid to flaccid intermediates takes place exactly at $\epsilon_1=-1$ (it is easy to show that the critical value of $\epsilon_1$ is independent of $\mu_0$ and $\beta$). Another example is the tree in  Fig. \ref{fig:Deg} with $N_P=6$ which shows the flaccid minimum energy state when minus $\epsilon_1$ exceeds a critical value. Transitions from rigid to flaccid as a function of $|\epsilon_1|$ become more common for increasing MLD. 

\section{Boltzmann Distribution.}

The minimum energy assembly pathways can be used to define ``low-temperature" \textit{Boltzmann Distributions} for the thirteen  probabilities $P_{n}$ that a tree molecule is associated with $n$ pentamers. The low-temperature limit applies if assembly and disassembly processes are entirely restricted to the minimum energy pathways discussed earlier. Under those conditions, the Boltzmann distribution equals
\begin{equation}
P_{n}\propto\exp\left(-\beta E(n) + n \ln(c_f)\right)
\label{BD}
\end{equation}
The last term in the argument of the exponential is obtained by replacing the reference chemical potential $\mu_0$ with the actual chemical potential $\mu=\mu_0+ \ln(c_f)$ where $c_f$ is the concentration of free pentamers in solution (in units of the reference concentration). The validity condition of the low-temperature approximation is that the quantities $\beta|\epsilon_i|$ are larger than one.

It can be checked that it follows from the Boltzmann distribution that
\begin{equation}
\frac{c_f^{12} r_0}{c_{12}}=K
\label{LMA}
\end{equation}
where $K=\exp\beta E(12)$, $r_0$ the concentration of free RNA molecules and where the $c_n$ are the concentrations of RNA molecules associated with $n$ pentamers. If $r_t$ is the total solution concentration of genome molecules then $P_n=c_n/r_t$. Equation \ref{LMA} has the form of the Law of Mass Action (LMA) for a chemical reaction in which twelve free pentamers in solution ``react" with a genome molecule not associated with pentamers, forming a viral particle with twelve pentamers. The low temperature Boltzmann distribution Eq. (\ref{BD}) is thus consistent with a chemical reaction picture. In the context of the LMA, the quantity $K$ is known as the \textit{dissociation constant} of the reaction. This dissociation constant depends on the energy parameters only through the total assembly energy so it is independent of the MLD and $N_P $ numbers. 

\subsection{Conservation Laws}
The probability distribution $P_{n}$ is related to conservation laws for the number of pentamers and genome molecules. Conservation of genome molecules requires that $r_0+\sum_{n=1}^{12} c_n = r_t$. This is assured if $\sum_{n=0}^{12} P_{n}^{eq} = 1$. The proportionality factor in the definition of $P_{n}$ is determined by this condition. Next, conservation of pentamer molecules is assured if $c_f=c_0 \Gamma\thinspace([P_n])$ where
 \begin{equation}
\Gamma\thinspace([P_n])  \equiv 1 - (D/12)\sum_{n=1}^{12} n P_{n} 
\label{eq:gammadef}
\end{equation}
with $D=12 r_t/ c_0$, $c_f$ the concentration of free pentamers and $c_0$ the total pentamer concentration. 

The quantity $D$ is the mixing ratio mentioned in Section I as an important thermodynamic control parameter. If $D=1$ then there are exactly enough pentamers to encapsidate all genome molecules so $D=1$ corresponds to the stoichiometric ratio. Because $c_f=c_0 \Gamma\thinspace([P_n])$ depends on the set of all $P_n$, the thirteen Boltzmann equations are in fact coupled and need to be solved self-consistently.

\subsection{Self-Assembly}

\begin{figure}[htbp]
\begin{center}
\includegraphics[width=3in]{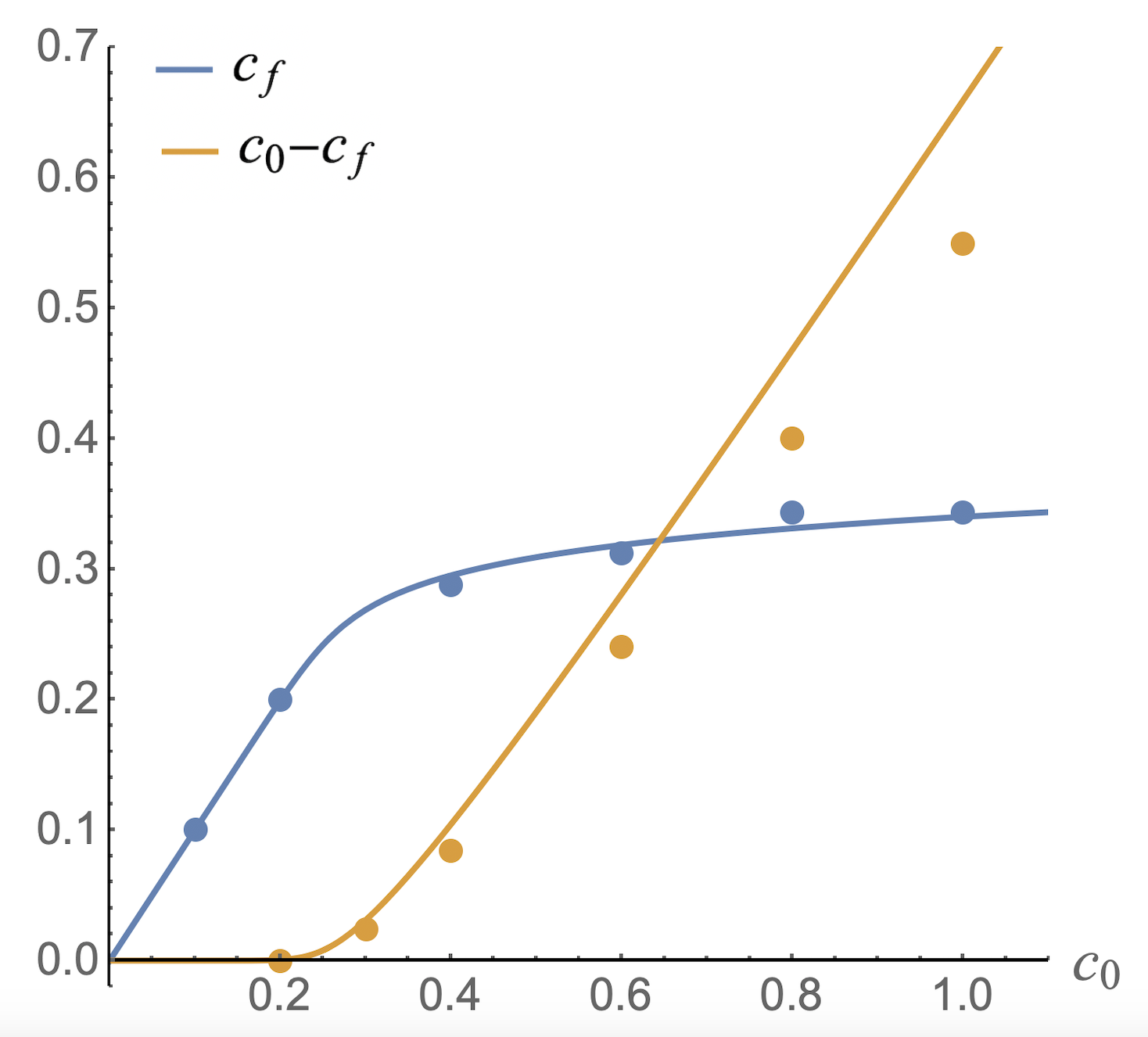}
\caption{Equilibrium self-assembly diagram for molecule (1) with the parameter values of Fig. \ref{example2}. Horizontal axis: total pentamer concentration $c_0$. Vertical axis: either the free pentamer concentration $c_f$ (blue) or the concentration $c_0-c_f$ of pentamers that are associated with a tree molecule (ochre). Solid lines: solution of Eq.\ref{B}}
\label{fig:EA}
\end{center}
\end{figure}

A standard diagnostic plot for self-assembling systems is that of the concentration of free building blocks and that of the assembled particles as a function of the total concentration of building blocks \cite{safran}. Such a plot was computed from the self-consistent solution of the Boltzmann distribution Eq.\ref{BD} for example molecule (1) shown in \ref{fig:dodtrees}. The energy parameters are the same as for Fig. \ref{example2}. Figure \ref{fig:EA} shows a plot of the concentrations of free pentamers in solution and that of pentamers that are part of an assembled particle as a function of the total pentamer concentration $c_0$.
For pentamer concentrations that are small compared to one (in our units), nearly all pentamers are free in solution so $c_f\simeq c_0$. As the total concentration increases, the concentration of free pentamers saturates when particle assembly starts. In this regime the concentration of pentamers that are part of an assembled particle increases roughly proportional to the total concentration. The resulting diagram is typical of that of self-assembling systems \cite{safran}.

The occupation probability $P_{12}$ for completed particles and the probability $P_{0}$ for pentamer-free genome molecules are, for these parameter values, large compared to the intermediate occupation probabilities with $1\leq n\leq11$.  If one can neglect these assembly intermediates then the conservation law for genome molecules reduce to $P_{0} \simeq(1-P_{12})$ and that of pentamers to $c_f\simeq c_0(1-DP_{12})$. Inserting these two relations into the LMA equation Eq.\ref{LMA} produces an (approximate) equation for the concentration $c_f$ of free pentamers:
\begin{equation}
\left(\frac{c_f}{c_0}\right)^{12}\left(\frac{D-1+\frac{c_f}{c_0}}{1-\frac{c_f}{c_0}}\right)\simeq\left(\frac{K}{c_0^{12}}\right)
\label{C}
\end{equation}
For $\left(\frac{c_0^{12}}{K}\right)$ small compared to one, this equation has a solution with $c_f$ close to $c_0$:
\begin{equation}
\frac{c_f}{c_0}\simeq 1 - \frac{D c_0^{12}}{K}
\end{equation}
For $\left(\frac{c_0^{12}}{K}\right)$ large compared to one and $D$ larger than one, the equation has a solution with $c_f$ independent of $c_0$:
\begin{equation}
{c_f}\simeq \left(\frac{K}{D-1}\right)^{1/12}
\end{equation}
Finally, for $\left(\frac{c_0^{12}}{K}\right)$ large compared to one but $D$ less than one, the equation has a solution with $c_f$ independent of $c_0$:
\begin{equation}
\frac{c_f}{c_0}\simeq 1 - D +\frac{K}{c_0^{12}}\frac{D}{(1-D)^{1/2}}
\end{equation}
There is thus a change in regimes when the depletion factor is equal to one. For the special case that $D=1$, the LMA equation reduces to
\begin{equation}
\left(\frac{c_f}{c_0}\right)^{13}\left(\frac{1}{1-\frac{c_f}{c_0}}\right)\simeq\left(\frac{K}{c_0^{12}}\right)
\label{B}
\end{equation}
The solid lines in Fig. \ref{fig:EA} for $K$ equal to $\exp \beta\Delta E$ with $\Delta E = 13.6$ are solutions of this equation. There is reasonable agreement with the concentrations computed from the full Boltzmann equation.

\subsection{Mixing Ratio.}

We can use the Boltzmann distribution to address the question how the mixing ratio $D$ affects assembly under equilibrium conditions. This is done by computing contours in the $c_0-D$ plane along which the packaging fraction $P_{12}$ is fixed. The dots in Fig. \ref{fig:PD} were obtained were computed numerically for two different values of $P_{12}$ for the same tree molecule and energy parameters as Fig. \ref{example2}:
\begin{figure}[htbp]
\begin{center}
\includegraphics[width=3.5in]{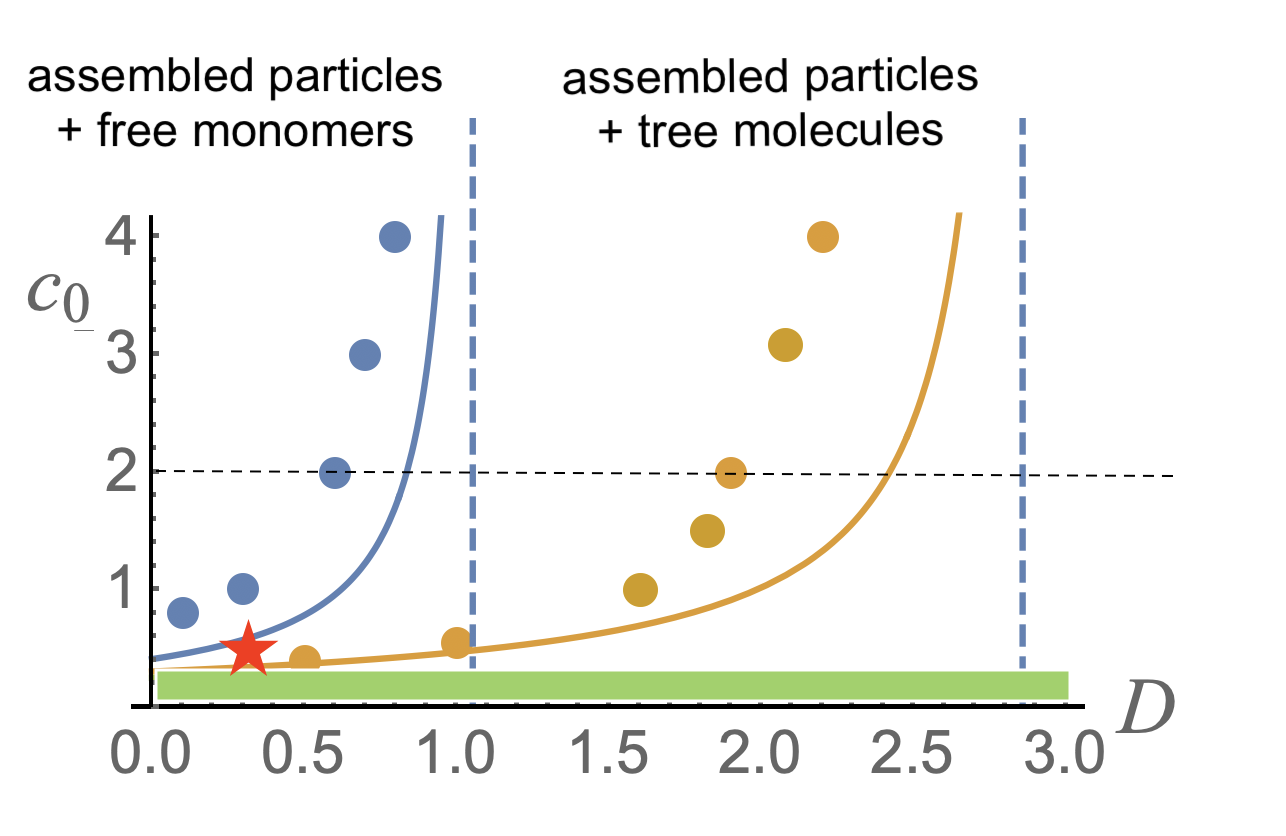}
\caption{Phase coexistence of assembled particles with excess free monomers and excess free tree molecules for $\epsilon_1=-0.2$, $\epsilon_3=-1$, and $\mu_0=-2$. Horizontal axis: Depletion factor $D$. Vertical axis: Pentamer concentration $c_0$. Blue dots: points where 95 percent of the genome molecules are packaged. Most of the remaining pentamers are free. Ochre dots: points where 35 percent of the tree molecules are packaged. The green sector is below the critical aggregation concentration at $c_0\simeq 0.2$ so there are no capsids in the green sectors. Solid blue and ochre lines: contours of fixed packaging fraction as computed from Eq.\ref{contour} under the neglect of assembly intermediates. Deviations between the solid lines and the dots are a measure of the importance of assembly intermediates. The red star is a region of the coexistence diagram where there is strong kinetic selection of genome molecules.}
\label{fig:PD}
\end{center}
\end{figure} 

Contours of fixed genome packing fraction can be computed analytically if one again neglects assembly intermediates. This approximation produces the family of hyperbolae
\begin{equation}
c_0(D)\simeq\frac{1}{(1-D\thinspace P_{12})}\left(\frac{K P_{12}}{1-P_{12}}\right)^{1/12}
\label{contour}
\end{equation}
Note that $c_0(D)$ diverges at $D=1/P_{12}$. The agreement between the approximate equilibrium theory and the actual values is reasonable for the $P_{12}=0.95$ contour (solid blue line vs. blue dots). The blue curve in Fig. \ref{fig:PD} can be viewed as an equilibrium phase boundary that separates two forms of phase coexistence. To the left, nearly all tree molecules are encapsidated with assembled particles in coexistence with excess free monomers while to the right most pentamers are part of assembled particles in coexistence with excess free tree molecules. This is consistent with the intuitive chemical reaction picture discussed in the introduction section. It seems surprising that the boundary line is shifted to values of $D$ below the stoichiometric ratio $D=1$ when $c_0$ is reduced. The reason for this shift becomes evident if one recalls there can be no assembly for $c_0$ less than the CAC for empty capsid assembly. According to Fig. \ref{fig:PD}, this CAC is around $0.2$ (green sector). The threshold value of $D$ below which the state of assembled particles dominates thus necessarily has to go to zero as $c_0$ approaches the CAC.   

\begin{figure}[htbp]
\begin{center}
\includegraphics[width=2.5in]{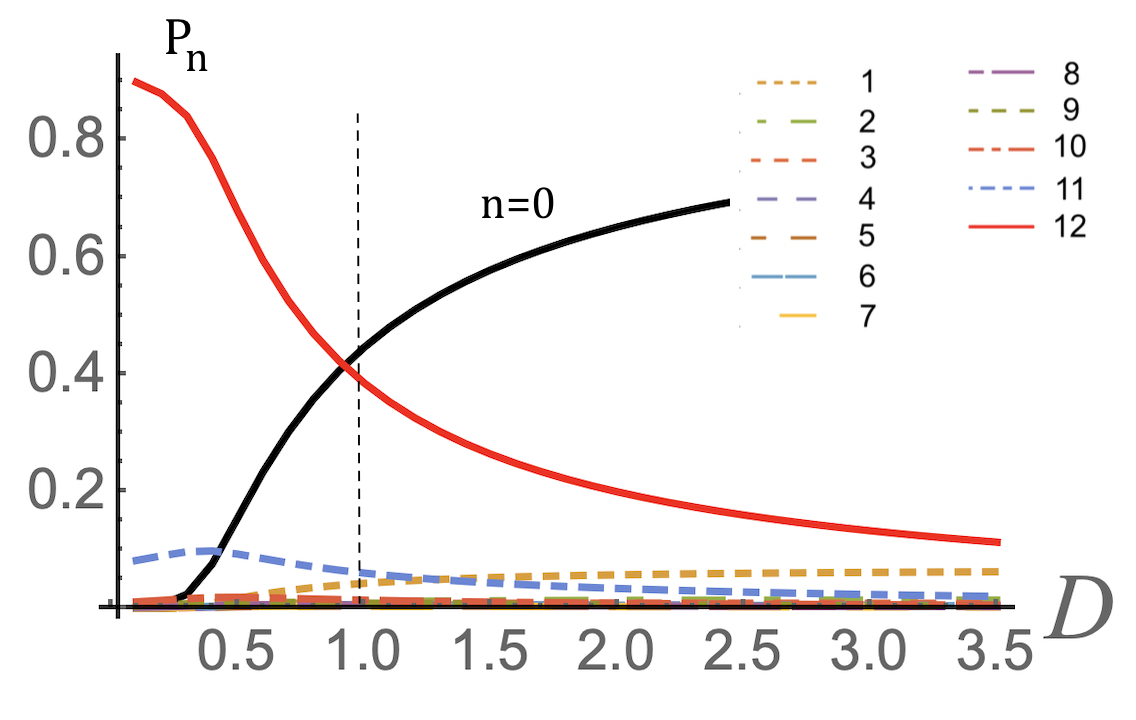}
\includegraphics[width=2.5in]{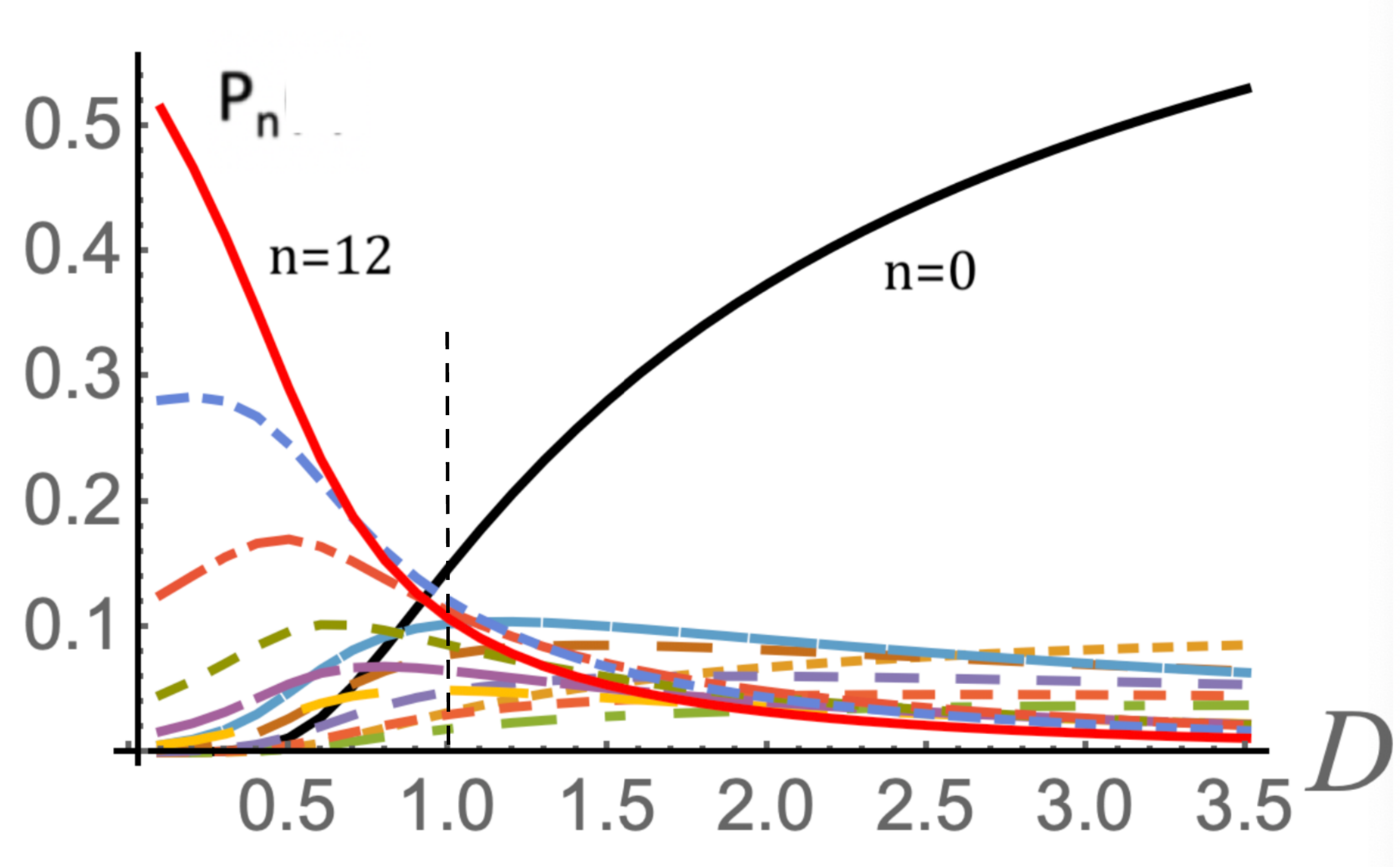}
\caption{Equilibrium occupation numbers as a function of the depletion factor $D$ for $c_0=1.0$. Top: same parameters as Fig. 10. The stoichiometric ratio D=1 is indicated. Bottom: same parameters as Fig.10 except that the ratio of the attractive interactions between the RNA-capsid and the capsid-capsid interactions has increased from $0.2$ to $1.2$. The reference chemical potential $\mu_0$ was reduced to $-5$ so the total assembly energy remains approximately the same. }
\label{fig:D}
\end{center}
\end{figure}

The agreement of the $P_{12}=0.35$ contour (ochre line) with the actual values (ochre dots) is poor. This can be traced to the neglect of assembly intermediates. Note also that the $P_{12}=0.35$ contour diverges at larger values of $D$ than the $P_{12}=0.95$ contour. Figure \ref{fig:D} (Top) shows that for larger values of $D$ the relative contribution of assembly intermediates indeed becomes comparable to the concentration of assembled capsids, which invalidates the assumption used to obtain Eq.\ref{contour}.  

When the ratio between the genome/pentamer and pentamer/pentamer interaction strengths is increased from $-\epsilon_1=0.2$ to $-\epsilon_1=1.2$ then the contribution from the partially assembled particles overwhelms that of fully assembled capsids for $D$ larger than one, as shown in Fig. \ref{fig:D} (bottom). This is due to the fact that when $D$ increases more and more pentamer binding sites become available as there are more tree molecules. Breaking up assembled particles and distributing the pentamers over the additional binding sites increases the entropy of the system while larger values of the binding energy of individual pentamers to specific edges of the tree molecules make up for the loss of pentamer-pentamer adhesion. As a result a high-entropy state with a distribution of partial shells has a lower free energy than the state of coexistence of viral particles with excess genome molecules. This transition can be viewed as a form of an \textit{order-disorder transition}. While such a transition would seem to conflict with the earlier chemical reaction picture, in actuality the LMA itself remains valid for Fig. \ref{fig:D}(Bottom) provided all assembly intermediates.

\section{Kinetics.}
In this section we complete the definition of the model by specifying the kinetics. We first consider the case where there is only one kind of spanning tree molecule in solution.

\subsection{Master Equation.}

The assembly kinetics of the model is defined in terms of thirteen occupation probabilities $P_n(t)$ that now depend on time. The assembly and disassembly dynamics for a given tree molecule with a particular assembly pathway is assumed to obey Markov chain statistics \cite{Perkett2014} for which the occupation probabilities evolve in time according to the master equation \cite{vanKampen}:
\begin{equation}
\frac{dP_n(t)}{dt}=\sum_{m= n\pm1}\left(W_{m,n}P_m(t)-W_{n,m}P_n(t)\right)
\label{eq:ME}
\end{equation}
Here, $W_{m,n}$ is a thirteen-by-thirteen matrix of transition rates from state $m$ to state $n$. We only include transitions with $m=n \pm1$ so with only one pentamer gained or lost at a time. Note that the diagonal entries of $W_{m,n}$ are not specified at this point. The matrix of transition rates $W_{m,n}$ will be defined in terms of simple diffusion-limited chemical kinetics (see Eq. 8.35 of ref.\cite{schulten}) where the addition of a pentamer to a genome molecule associated with a pentamer cluster of size $n$ is treated as a bimolecular reaction with a rate $k_{n,n+1} {r_{n}}{c_f}$ where $k_{n,n+1}$ is the on-rate defined as 
\begin{equation}
 k_{n,n+1}=\lambda
 \begin{cases}
 e^{-\beta(E(n+1)-E(n))}\quad\quad\thinspace\thinspace \text{if}\thinspace \quad E(n+1) > E(n)\\
1\qquad\qquad\qquad\quad\qquad\thinspace \text{if} \quad E(n+1) < E(n)
 \end{cases}
 \end{equation}
Here, $\lambda$ is a base rate that depends on quantities like the diffusion coefficient but that is independent of concentration. If adding a pentamer reduces the energy then the rate is equal to the base rate. If there is an energy cost to adding a pentamers, then the base rate is reduced by an Arrhenius factor, similar to the Metropolis algorithm of Monte-Carlo simulations. 

The on-rates are related to the rate matrix by $W_{n,n+1}=k_{n,n+1}{c_f}$. The entries of the rate matrix that correspond to adding a pentamer are then
 \begin{equation}
 W_{n,n+1}=\lambda c_0 \Gamma\thinspace([P_i])
 \begin{cases}
e^{-\beta\Delta E_{n,n+1}}\qquad\text{ if   } \Delta E_{n,n+1}>0\\
1 \qquad\qquad\qquad\thickspace\thinspace\text{ if   }  \Delta E_{n,n+1}<0
 \end{cases}
 \end{equation}
where $\Delta E_{n,n+1} \equiv (E(n+1)-E(n))$. The off-rate entries $W_{n+1,n}$ are determined by the on-rates through the condition of \textit{detailed balance}:
\begin{equation}
\frac{W_{n+1,n}}{W_{n,n+1}}=\frac{P_{n}}{P_{n+1}}
\end{equation}
where on the right hand side the Boltzmann Distribution Eq. (\ref{BD}) must be inserted. This results in 
 \begin{equation}
 \begin{split}
 &W_{n,n+1}=c_0 \Gamma\thinspace([P_i])
 \begin{cases}
e^{-\beta\Delta E_{n,n+1}}\qquad \Delta E_{n,n+1}>0\\
1 \qquad\qquad\qquad\thickspace  \Delta E_{n,n+1}<0
 \end{cases}
 \\&W_{n+1,n}=\qquad\qquad\thickspace 
 \begin{cases}
1\qquad\qquad\qquad\thickspace  \Delta E_{n,n+1}>0\\
e^{\beta\Delta E_{n,n+1}}\qquad\thickspace\thickspace   \Delta E_{n,n+1}<0
 \end{cases}
 \end{split}
 \end{equation}
where the base rate $\lambda$ has been absorbed in a redefinition of time. 
 Note that the master equation is nonlinear because of the dependence of the factor $\Gamma(|P_i|)$, as defined in Eq.  (\ref{eq:gammadef}), on the occupation probabilities. 
 
\begin{figure}[htbp]
\begin{center}
\includegraphics[width=3.3in]{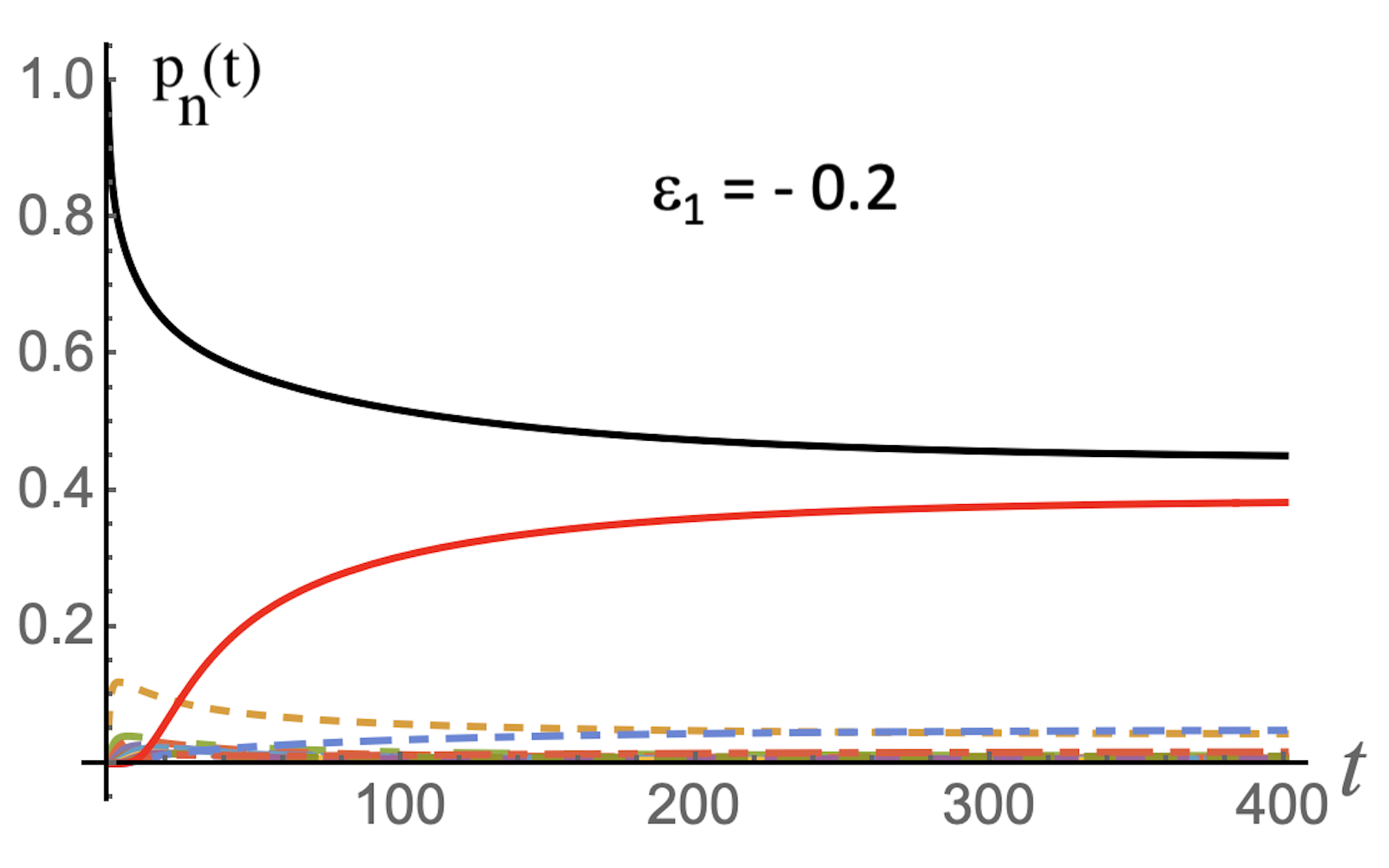}
\includegraphics[width=3.3in]{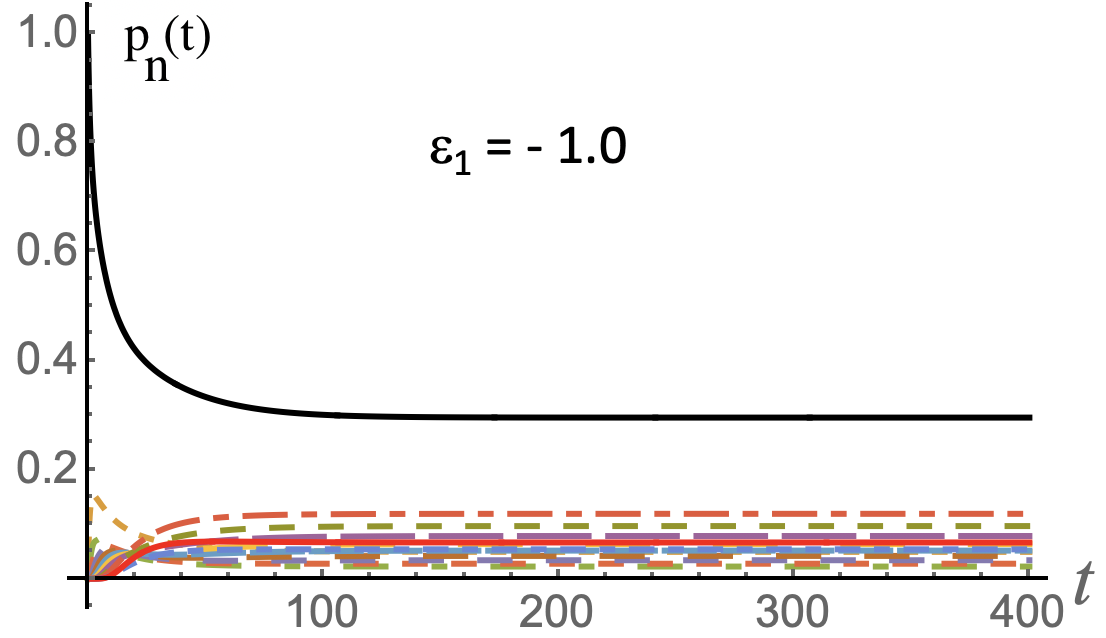}
\caption{Numerical solution of the master equation for molecule (1). Top: parameter values are those of Fig. \ref{example2} with $\epsilon_1=-0.2$. Bottom: same except that $\epsilon_1=-1.0$. Shown are the occupation probabilities $P_n(t)$. The color code is the same as that of Fig.14. Time is in dimensionless units of $1/\lambda$.}
\label{ME}
\end{center}
\end{figure}

Some examples of numerical solutions of the master equation are shown in Fig. \ref{ME} for molecule (1) under reference conditions $c_0=D=\beta=1$. 
The top figure shows the case of $\epsilon_1=-0.2$. Other energy parameters are those of Fig. \ref{fig:EA}), which shows the energy profile that corresponds to the figure. The red curve is the probability $P_{12}(t)$ for a tree to be encapsidated, the black curve the probability $P_0(t)$ for a tree to be free of pentamers. Eventually, about forty percent of the genome molecules are encapsidated by complete capsids with $n=12$ pentamers while about five percent are encapsidated by partial capsids with $n$ less than 12. We checked the assembly pathway and it is consistent with the Zlotnick Model shown in Fig. \ref{fig:Zlotnick Model}. The occupation probabilities exponentially approach constant values at late times that agree with the Boltzmann distribution. Closely similar results are obtained if one replaces molecule (1) by molecule (2). At the this point, the packaging kinetics does not seem to be able to really distinguish between the two molecules.

The bottom figure shows the effect of increasing the relative strength of the interaction between pentamers and genome molecules to $\epsilon_1=-1.0$, while the reference chemical potential is reduced to $\mu_0=-5.2$ for the reason discussed above Fig. \ref{fig:EA}. The final equilibrium state is now a polydisperse mixture of aggregates of various sizes, consistent with Fig:\ref{fig:D}. While for $\epsilon_1=-1.0$, the assembly pathway still follows that of the Zlotnick Model for both molecule (1) and (2), for $\epsilon_1=-1.2$ the assembly pathway of molecule (2) is completely changed. Now flaccid assembly intermediates appear, such as the shown in Fig. \ref{fig:F}.

\subsection{Time Scales}
\begin{figure}[htbp]
\begin{center}
\includegraphics[width=3.5in]{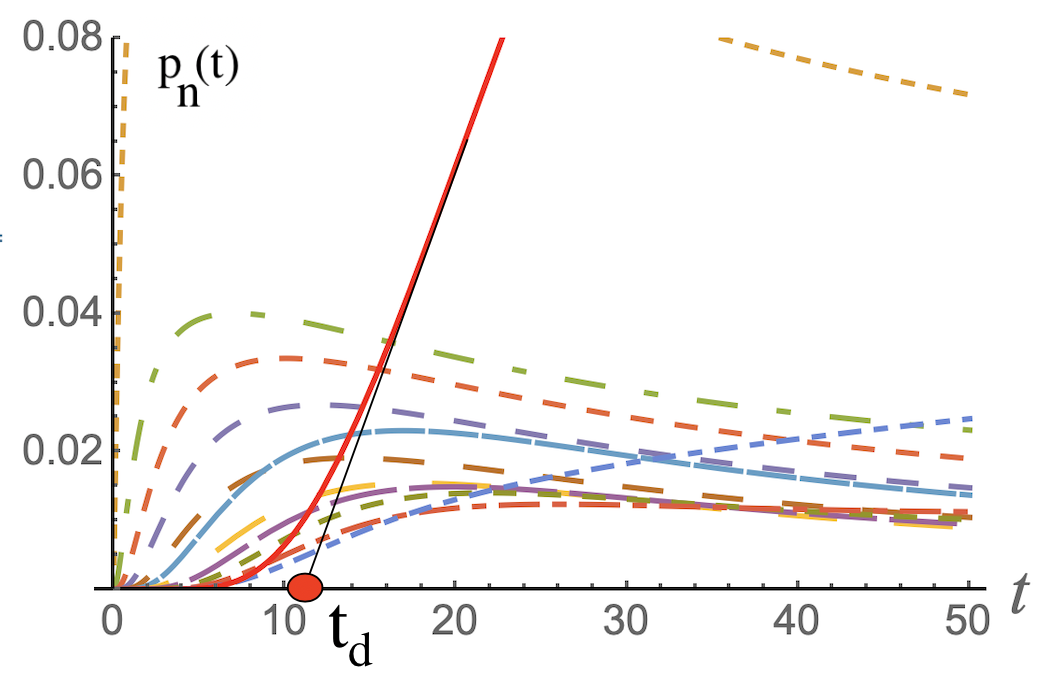}
\caption{Assembly shock-wave for molecule (1) for $\epsilon=-0.2$. The intersection of the maximum tangent of the $P_{12}(t)$ curve with the time axis (solid black line) defines the assembly lag time $t_d$.}
\label{fig:master3}
\end{center}
\end{figure}

The occupation probabilities in Fig. \ref{ME} are governed by multiple scales. The first important time scale is defined in Fig. \ref{fig:master3}, which shows the early stages of the assembly of molecule (1) for $\epsilon=-0.2$ but now in magnified form as compared with Fig.\ref{ME}(top).
The \textit{delay time} $t_d$ is defined as the intercept of the maximum tangent of $P_{12}(t)$ with the time axis. This gives $t_d\simeq11$ for the case (of molecule (1) with $\epsilon_1=-0.2$. As mentioned in Section I, assembly delay times are a familiar feature of the assembly kinetics of empty capsids \cite{Casini2004} and of aggregation phenomena in general \cite{Wu1992}. A second important time scale refers to the late-time relaxation of the occupation probabilities towards the Boltzmann distribution. Suppose one completes the definition of the transition matrix by introducing the diagonal entries $W_{n,n}=-\sum_{m\neq n}W(m,n)$. The resulting matrix $W_{m,n}$ now has column elements adding to zero. Using this completed transition matrix, the master equation can be rewritten in the form of a matrix equation $\frac{d\bf{P}}{dt}=\bf{W P}$. It immediately follows that eigenvalues of the completed transition matrix are the decay rates of the various modes that correspond to the eigenvectors. The relaxation rate determining the approach to final equilibrium is the smallest eigenvalue of $W_{m,n}$. The \textit{relaxation time} $t_r$ is defined as the inverse of the smallest eigenvalue so $P_n(t)-P_0(\infty)\propto \exp-t/t_r$ in the late time limit for any $n$. Figure \ref{ME} shows that this time-scale is much longer than the delay time $t_d$, specifically $t_r\simeq644$ so about two orders of magnitude longer than $t_d$. The different functions $P_n(t)$ in Fig. \ref{fig:master3} display a maximum as a function of $n$. The corresponding peak times increase with $n$, thus describing a type of assembly shock-wave propagating in configuration space from small to large $n$. Similar assembly shock waves have been reported for dynamical versions of the Zlotnick Model \cite{Endres2002, Morozov2009}. The dependence of the two time scales on the energy scale is also quite different. For $\beta=3$, $t_d\simeq11.7$ while $t_r\simeq1.7\times 10^5$ producing a significant separation in time-scales. Relaxation to equilibrium is evidently an activated process, which is not surprising given the form of the assembly energy profiles. In addition, increasing $\beta$ from one to three also has the effect of suppressing the partially assembled capsids in Fig. \ref{ME} (top). This remains the case if the mixing ratio $D$ is increased from one to two, which now produces simple phase coexistence of fully assembled particles and bare genome molecules.

\begin{figure}[htbp]
\begin{center}
\includegraphics[width=3.5in]{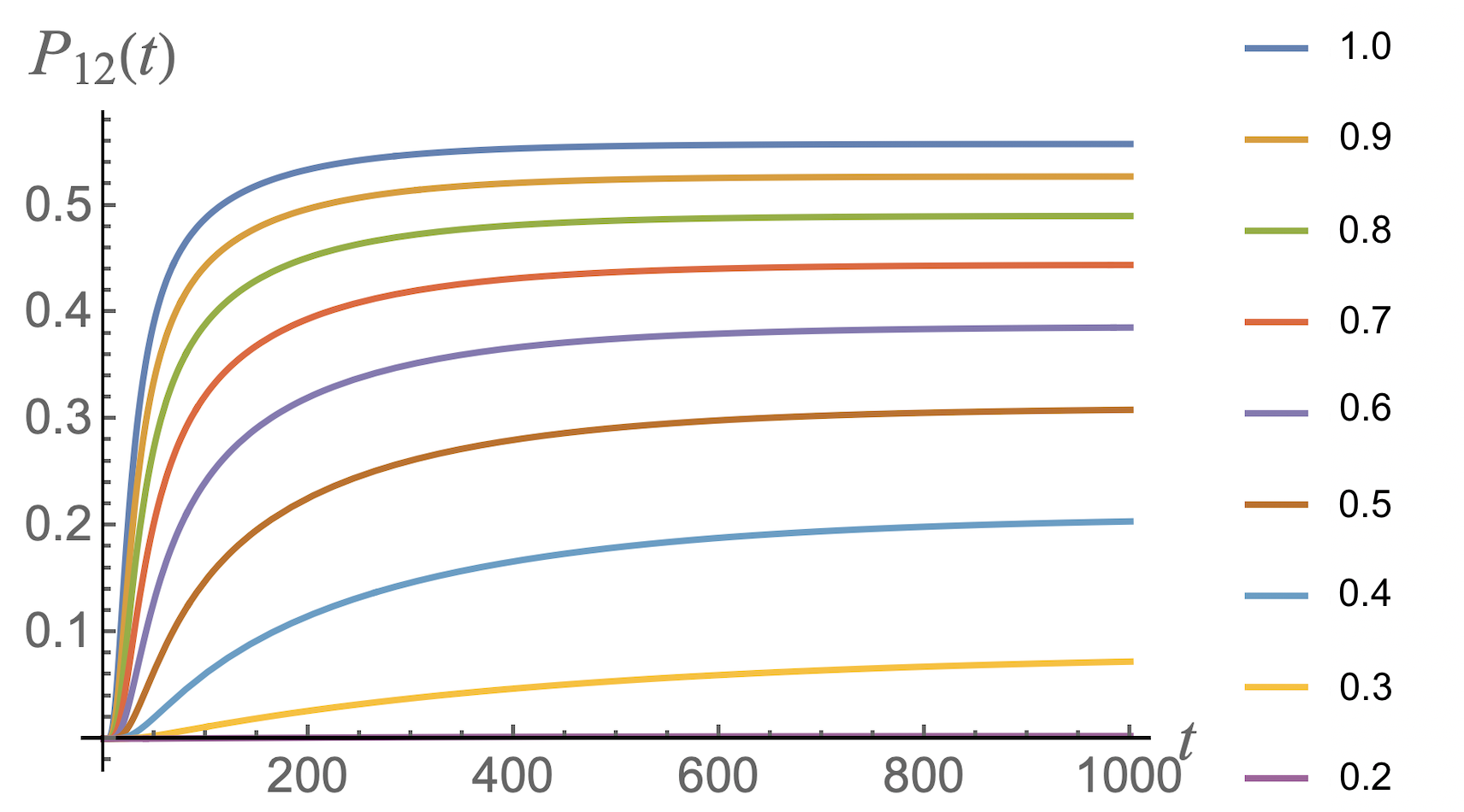}
\caption{Plots of the fraction $P_{12}(t)$ of genome molecules that are fully encapsidated for increasing total concentration for different total pentamer concentration $c_0$ ranging from 1.0 to 0.2. Particles do not form for $c_0$ less than 0.2.}
\label{LS}
\end{center}
\end{figure}

Figure \ref{LS} shows the effect on the kinetics of varying the total pentamer concentration $c_0$. 
The plot qualitatively reproduces the time-dependent light scattering studies of the assembly of empty capsids \cite{Casini2004}. Note that there is no assembly if the pentamer concentration drops below about $0.2$, which is close to the CAC shown in Figs \ref{fig:EA} and \ref{fig:PD}.  Fixing the concentration and varying the depletion factor produces similar-looking plots.

\subsection{Disassembly Kinetics.}

\begin{figure}[htbp]
\begin{center}
\includegraphics[width=2.7in]{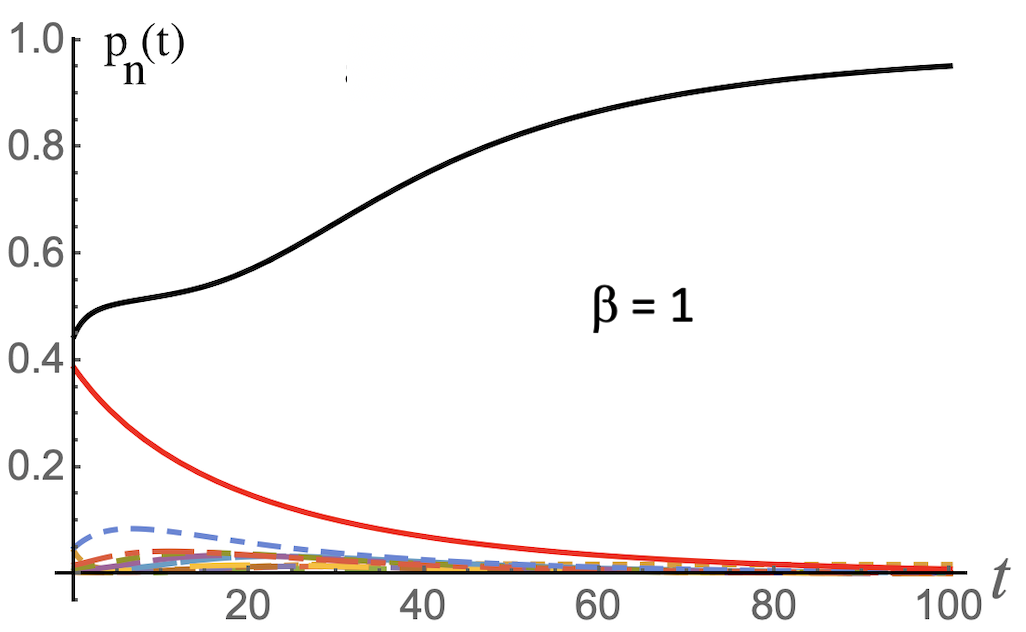}
\includegraphics[width=2.7in]{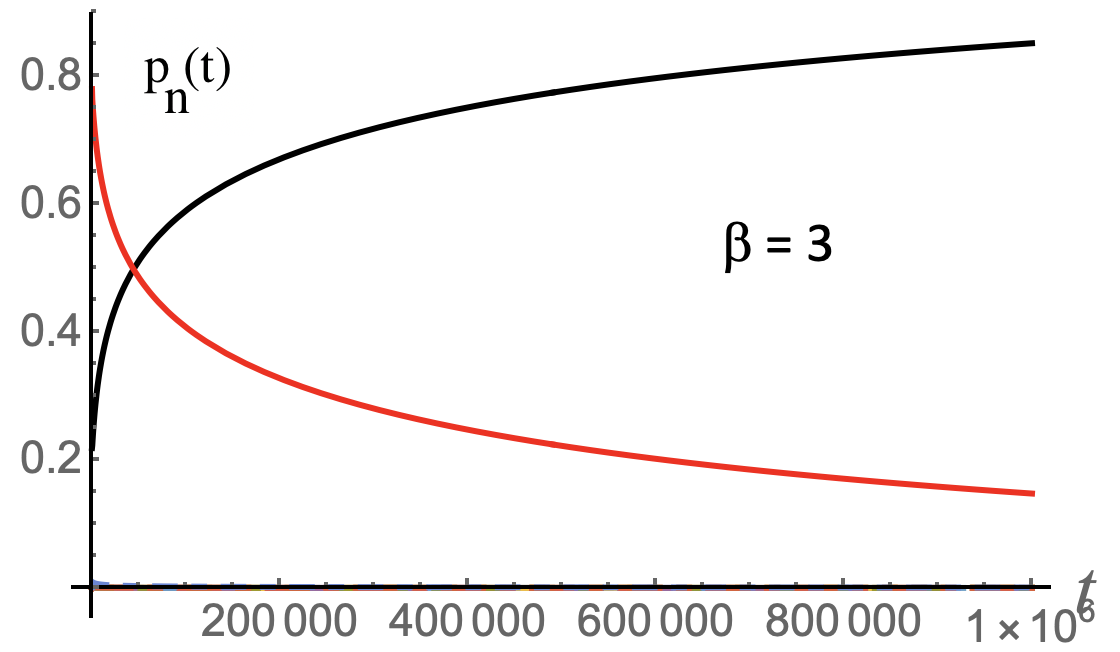}
\caption{Top: The assembly run of Fig. \ref{fig:D} at $c_0=1$ and $\beta=1$ followed by a disassembly run at $c_0=0.1$. The color coding is that of Fig.12. The assembled particles evaporate. Bottom: Same as the top figure except that the temperature is reduced to $\beta=3$. }
\label{fig:dis}
\end{center}
\end{figure}

An important question concerns the fate of assembled particles when the solution concentration $c_0$ of pentamers is reduced. This can be probed by using the outcome of an assembly run with the total pentamer concentration $c_0=1$ as the initial condition for a second run with a ten-fold reduced pentamer concentration, so with $c_0=1$ reduced to $c_0=0.1$. The results are shown in Fig. \ref{fig:dis}, still for molecule (1).
For $\beta=1$, the assembled particles rapidly disintegrate when the pentamer concentration is reduced. The time-scale for the disintegration is of the order of the relaxation time $t_r\simeq 31.6$, which is reduced compared to the case of $c_0=1$. The energy activation barrier is, in this case, unable to ``protect" the assembled particle from disintegration by thermal activation. Recall we found that when $\beta$ was increased from $1$ to $3$, then the delay time $t_d$ barely increased while the relaxation time $t_r$ increased by three orders of magnitude. Repeating the assembly-disassembly run for $\beta=3$ (bottom figure), produced the result that the concentration of assembled particles still decreased with time during the disassembly run but now on time-scales of the order of $10^6$. In contrast, during the assembly run about fifty percent of the genome molecule had been encapsidated by a time $t\simeq10^4$. This assembly time scale is significantly longer than the delay time but still is two orders of magnitude shorter than the characteristic time scale for disassembly. We attribute this to the fact the activation energy barrier for assembly is significantly lower than for disassembly for a supersaturated system. Such a separation in assembly and disassembly time scales would seem to be an essential condition for a functioning viral particle. Below we will focus on the case of $\beta=3$.

Next, we carried out assembly-disassembly runs for $\epsilon_1=-1.2$ and $\mu_0=-5.6$. The assembled particles and the various intermediate aggregates rapidly ``evaporated" during the disassembly run even for $\beta=3$. The reason for the loss of stability can be seen by comparing the minimum energy assembly profiles for $\epsilon_1=-0.2$ and $\epsilon_1=-1.2$ if the reference chemical potential is set equal to $\mu^*$.
\begin{figure}[htbp]
\begin{center}
\includegraphics[width=2.4in]{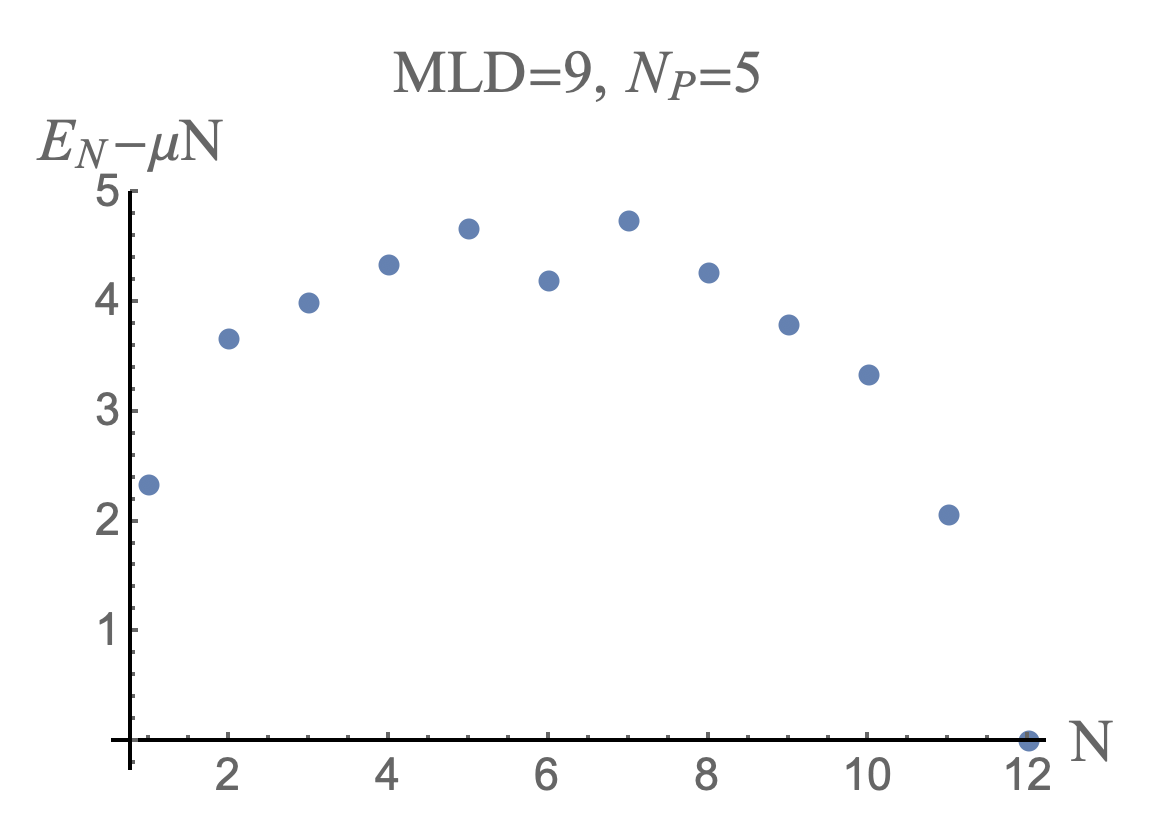}
\includegraphics[width=2.4in]{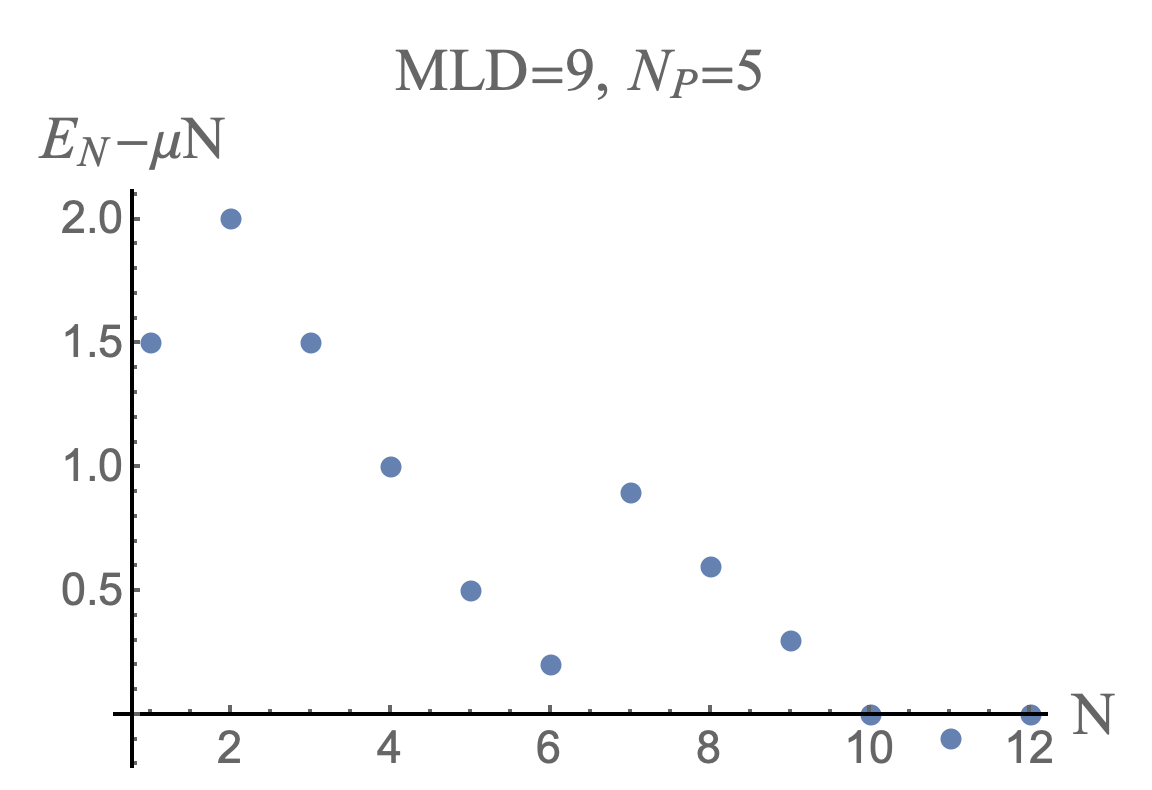}
\caption{Assembly energy profiles for $\epsilon_1=-0.2$ (top) and $\epsilon_1=-1.2$ (bottom) under conditions of assembly equilibrium with $\mu_0=\mu^*$.}
\label{fig:dis3}
\end{center}
\end{figure}
For $\epsilon_1=-0.2$, the energy profile is quite similar to that of the Zlotnick Model (see Fig. \ref{fig.2}). A standard nucleation and growth scenario is expected for the assembly process. On the other hand, $E(n)$ has a deep minimum at $n=6$ for $\epsilon_1=-1.2$ under assembly equilibrium conditions. This corresponds to half-assembled particles, which is the \textit{exact opposite} of what is predicted by the nucleation and growth scenario. Recall that in that case $n=6$ is expected to be close to an energy maximum. Increasing the magnitude of $\epsilon_1$ indeed progressively destroys the activation energy barrier that ``protects" the assembled state. 

\section{Packaging Competition}

\begin{figure}[htbp]
\begin{center}
\includegraphics[width=3.5in]{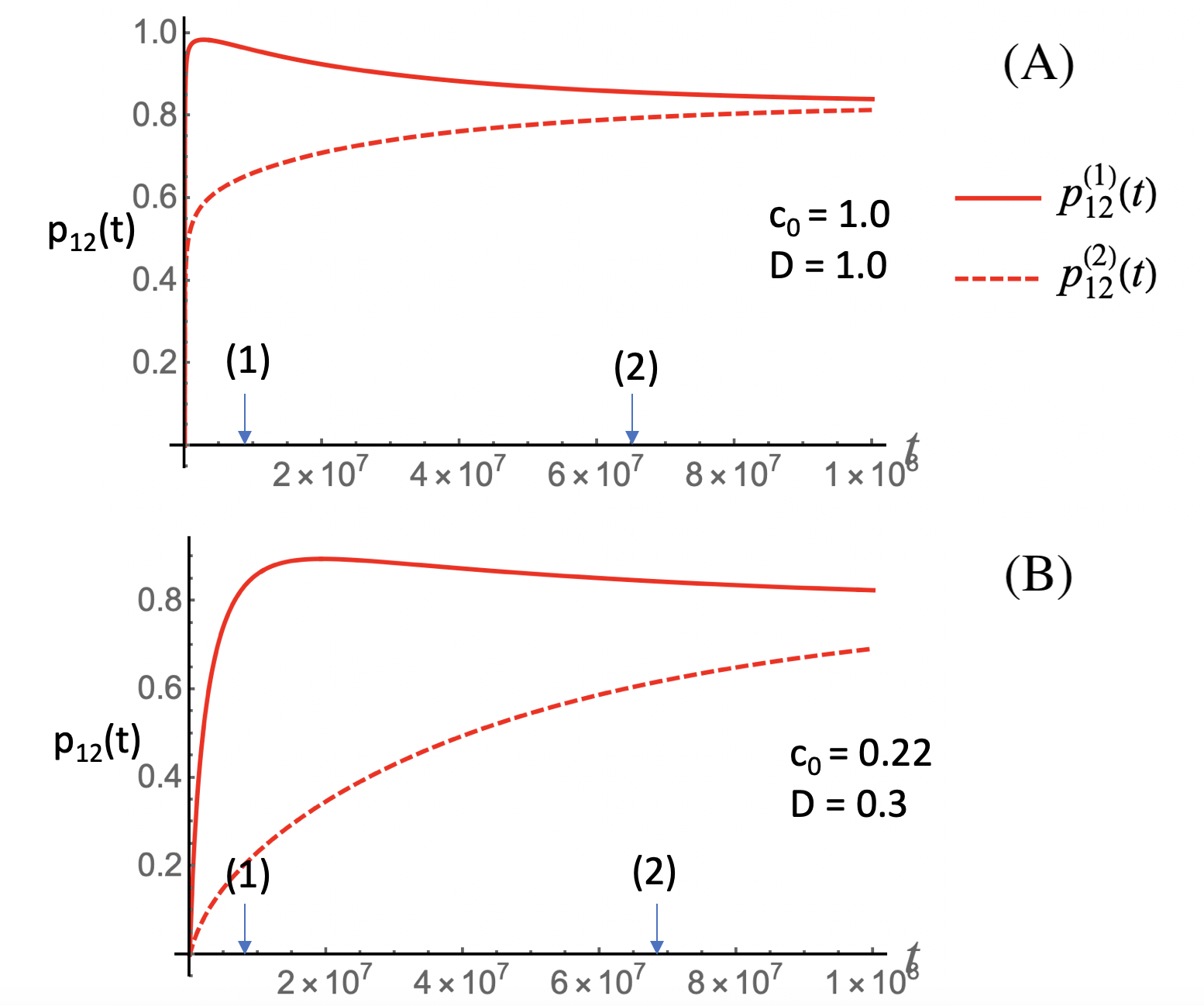}
\caption{Time-dependence of the fraction of packaged tree molecules of example molecules (1) and (2). (A): $\beta=3$, $c_0=D=1$ and $\mu_0=-2.5$. The relaxation times of the two molecules are indicated by the two arrows. (B): Same as (A) except that $c_0=0.22$ and $D=0.3$.}
\label{RS}
\end{center}
\end{figure}

We are now in a position to carry out packaging competitions between two genome molecules that either have a different tree topology or that have the same tree topology but different dodecahedral wrapping configurations.  We did this for a system containing a concentration $c_0$ of pentamers as well as equal concentrations of molecules (1) and (2) that competed with each other for packaging for the case $\epsilon_1=-0.2$. 
Let $P_n^{(1)}(t)$ and $P_n^{(2)}(t)$ be the corresponding sets of occupation probabilities.  Conservation of both types of tree molecules requires that the occupation probabilities $P_n^{(1)}(t)$ and $P_n^{(2)}(t)$ separately sum to one. Conservation of pentamers is satisfied if $(c_f/c_0)=\Gamma\thinspace([P_n])$ obeys the condition
\begin{equation}
\Gamma\thinspace([P_n])=1 - (D/24)\sum_{n=1}^{12} n (P_n^{(1)} + P_n^{(2)})
\label{cf}
\end{equation}
The two occupation probabilities $P_n^{(1)}(t)$ and $P_n^{(2)}(t)$ obey separate master equations of the form of Eq. (\ref{eq:ME}). The only difference is that now the factor $\Gamma\thinspace([P_n])$ in the expression for the transition matrix Eq.\ref{cf} couples the two master equations. Figure \ref{RS} shows an example of the solution of the two coupled master equations for two different parameter sets. 
The relaxation times are indicated by arrows. Based on the discussion in Section 2, it is expected that the highly branched molecule (1) will outcompete linear molecule (2) since it can accommodate more optimally wrapped pentamers during assembly. Recall however that in actuality the assembly energy profiles of the two molecules are not very different (see Fig.9). The maximum selectivity in Fig. \ref{RS}(A) of molecule (1) over molecule (2) for parameter values $\beta=3$, $c_0=D=1$ and $\mu_0=-2.5$. It is achieved very early in the assembly process. In order to be quantitative, define the relative packaging selectivity $S(1|2)$ of molecule (1) with respect to molecule (2) to be 
\begin{equation}
S(1|2)\equiv (P_{12}^{(1)}-P_{12}^{(2)})/(P_{12}^{(1)}+P_{12}^{(2)})
\label{S12}
\end{equation}
which ranges from one to minus one. This fraction depends on the time instant when probabilities are being computed. The maximum value of the relative packaging selectivity will be denoted by $S_m(1|2)$. For \ref{RS}(A), the maximum selectivity $S_m(1|2)\simeq0.43$. The selectivity $S(1|2)(t)$ then slowly decays to zero on a time scale of the order of the relaxation time $t_r^{(2)}$. The fact that the selectivity goes to zero in the long time limit can be understood from the fact that the total assembly energies of molecules (1) and (2) are the same, along with the fact that there are practically no intermediate assemblies. Selectivity is a purely kinetic effect.

To further increase the selectivity, we varied the total pentamer concentration $c_0$. For given $c_0$, the mixing ratio $D$ was given the maximum value for which at least 90 percent of the type (1) genome molecules were still encapsidated. The maximum selectivity was obtained when $c_0$ was reduced to a value close to the lower bound of $0.2$ (the CAC). Figure \ref{RS}(B) shows the case $c_0\simeq 0.22$ with $D=0.3$. The maximum selectivity is $S_m(1|2)\simeq0.8$. This is the highest selectivity that we were able to achieve. This increase in selectivity near the CAC is not the result of increasing relaxation times (see Fig. \ref{RS}(B)). If $c_0$ drops below the CAC then the selectivity of the particles that assemble remains about $0.8$ but only a very small fraction of the genome molecules are encapsidated. 

The fact that the kinetic selectivity is largest in the limit of small $c_0$ is intriguing. In an Arrhenius description the nucleation rate for particle assembly is the product of an attempt frequency and a Boltzmann factor $\exp-\beta\Delta E$ with $\Delta E$ the assembly activation barrier. The assembly energy profiles do not depend on the pentamer concentration $c_0$ so this increase of the selectivity when $c_0$ is reduced must be due to a dependence of the attempt frequency on $c_0$ in a manner that differentiates between the two molecules. According to Fig. \ref{example2}, the first two assembly steps of molecules (1) and (2) that bring the process to the energy maximum at $n=2$ are the same. Once past $n=2$, the negative slope of the energy profile for molecule (1) is significantly steeper than that of molecule (2). Only after three additional assembly steps does the energy profile of molecule (2) acquire a similarly large negative slope. This distinction is a consequence of the fact that the wrapping number of molecule (1) is six while it is two for molecule (2). The energy of capsid assemblies with $n$ greater than two thus decreases more rapidly for molecule (1). Moreover, the two maximally wrapped pentamers of molecule (2) are not adjacent. If the pentamer concentration is low then the packaging of molecule (2) will be significantly delayed as compared to molecule (1). The system performs a weakly directed random walk along the flattish top of the energy barrier with a significant probability for \textit{disassembly} of the assembly nucleus. This happens if the random walk returns to n=2. As a result, the effective attempt frequency of molecule (2) is significantly reduced. The reason why in the limit of small $c_0$ the mixing ratio $D$ must be reduced to well below stoichiometric ratio $D=1$ in order to achieve a 90 percent packaging probability is a consequence of the equilibrium thermodynamics discussed earlier. Note that the point $c_0(D=0)$ on the contour in the equilibrium phase-coexistence diagram Fig. \ref{fig:PD} for fixed 90 percent packaging probability is also the point with the lowest value for $c_0$.

\begin{figure}[htbp]
\begin{center}
\includegraphics[width=3.5in]{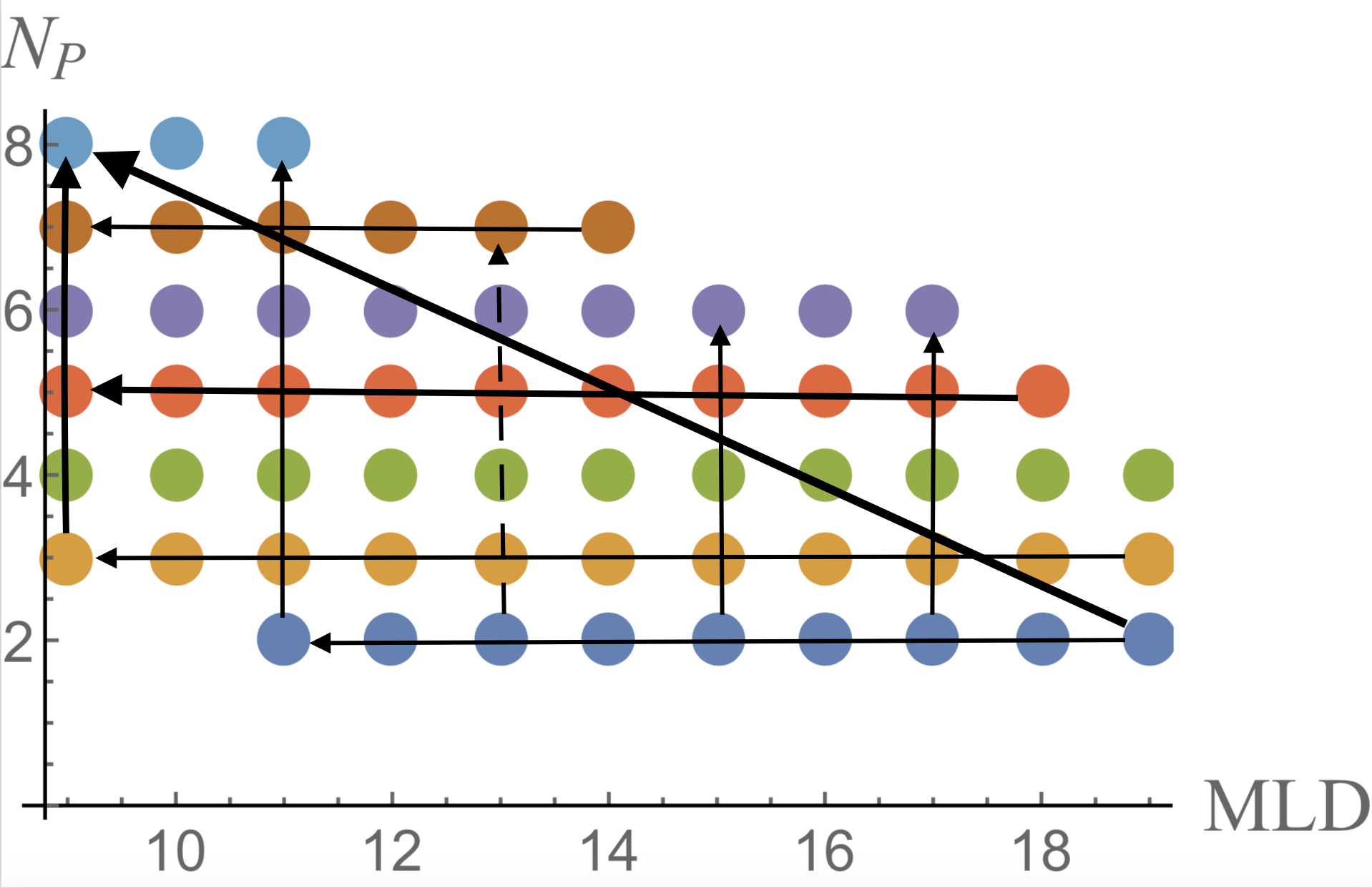}
\caption{Plot of wrapping numbers ($N_P$) and maximum ladder distances (MLD showing the outcomes of packaging contests indicated by arrows. Each arrow indicates the outcome of a contest, with the arrow pointing from the loser to the the winner. The width of the arrow is a measure of the maximum selectivity $S_m$. A dashed arrow indicates a weak selectivity. The energy parameters were $\beta=3$, $c_0=0.3$, $D=0.5$, $\epsilon_1=-0.2$, $\epsilon_3=-1$, and $\mu_0=-2.5$}
\label{PC2}
\end{center}
\end{figure}

We constructed a table of packaging selectivities of pairs of trees with different MLD and wrapping numbers (Fig. \ref{PC2}). This was done for $\beta=3$ and $c_0=0.3$ instead of $0.2$, which produced an increase of the mixing ratio for 90 percent packaging from $D=0.3$ to a more reasonable $D=0.5$. 
The result is clear: when two trees compete with the same wrapping number but different MLD, then the tree with the smaller MLD outcompetes the tree with the larger MLD. When two trees with the same MLD but different wrapping number compete, then the tree with the larger wrapping number outcompetes the tree with the smaller wrapping number. 

\begin{figure}
     \centering
         \includegraphics[width=3.in]{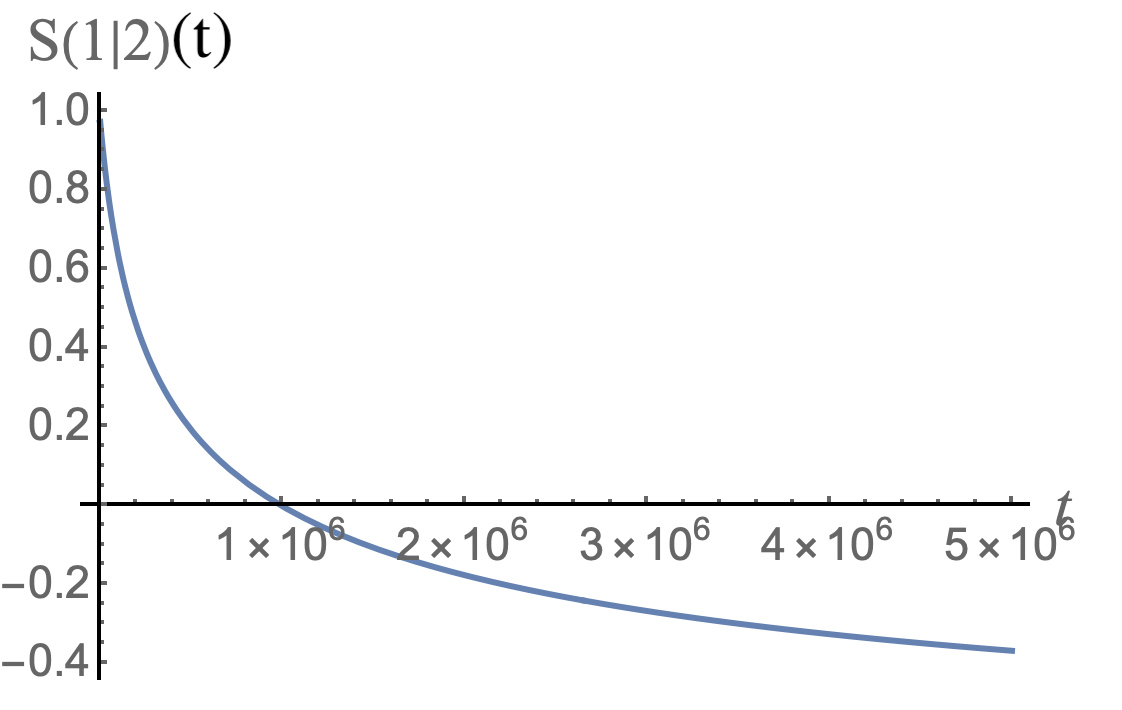}
        \caption{Time-dependence of the packaging selectivity $S(1|2)(t)$ of molecule (1) with respect to molecule (2). The energy parameters are $\epsilon_1=-1.0$, $\epsilon=-1$, and $\mu_0=-5.2$}
        \label{fig:PC}
\end{figure}

It would seem that one should be able to achieve an even higher selectivity by increasing $|\epsilon_1|$, the ratio of strength of the genome-pentamer attraction and the pentamer-pentamer attraction.  The outcome of a packaging contest between molecules (1) and (2) is shown in Fig. \ref{fig:PC} for the case $|\epsilon_1|=1$. The reference chemical potential was adjusted so the energy activation barrier was the same as for our standard value $|\epsilon_1|=0.2$.
Figure \ref{fig:PC} shows the time-dependence of the selectivity $S(1|2)$. During the first stage of the assembly---with $t$ less than $10^4$---$S(1|2)$ rises very quickly to (nearly) one, which agrees with the expectation that increasing $|\epsilon_1|$ should enhance the packaging selectivity. However, as time progresses, fully packaged type (1) molecules particles start to disassemble while additional molecule (2) assemblies start to appear. The packaging selectivity drops, goes to zero and then \textit{changes sign}. This ``inversion"  is in fact a generic feature of packaging competitions for large values of $|\epsilon_1|$. Eventually, about 70 percent of the type (2) molecules are packaged in complete capsids. A large fraction of type (1) molecule are associated with a polydisperse distribution of incomplete particles. Recall from Section IV that increasing $|\epsilon_1|$  produces a large fraction of \textit{incomplete particles}. Recall also that for large $|\epsilon_1|$ assembled particles easily disintegrate when the pentamer concentration is reduced.  

\begin{figure}
     \centering
         \includegraphics[width=3.4in]{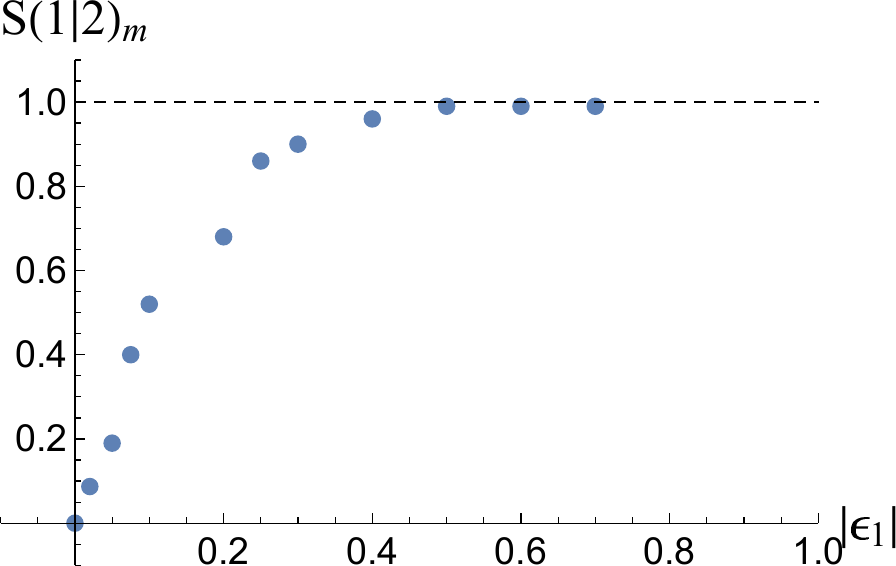}
        \caption{Maximum packaging selectivity $S(1|2)_m$ of molecule (1) with respect to molecule (2) as a function of $|\epsilon_1|$. The reference chemical potential $\mu_0$ was adjusted to maintain the activation energy barrier at an approximately constant value.}
        \label{MPS}
\end{figure}

Since increasing $|\epsilon_1|$ leads to thermodynamically unstable particles, how small can one make $|\epsilon_1|$ and still have a reasonable selectivity? Figure \ref{MPS} shows how the maximum packaging selectivity $S(1|2)_m$ depends on $|\epsilon_1|$ in the regime of small $|\epsilon_1|$.
The maximum packaging selectivity rises very steeply from zero for increasing $|\epsilon_1|$ and reaches it maximum value around $|\epsilon_1|\simeq0.4$. Polydispersity and selectivity inversion only appears for larger values of $|\epsilon_1|$.

\section{Conclusion}

We have presented a simple, quasi-analytical model for the study of the packaging of tree-like genome molecules inside dodecahedral capsids composed of twelve pentamers. The model is sufficiently simple that the kinetics can be obtained from the numerical solution of a set of coupled master equations. The genome molecules are represented by a ``core" of specific links with enhanced affinity for the edges and that visits all vertices of the dodecahedron but covers only nineteen of the edges. The spanning tree is complemented by eleven non-specific links with reduced edge affinity. The minimum energy assembly pathways are characterized by two numbers: the maximum ladder distance (MLD) and the wrapping number ($N_P$.) The MLD of a spanning tree is a topological invariant of the tree that is a global measure of the ``branchiness" of a tree while $N_P$ is a geometrical measure of the surface distribution of specific links over the dodecahedron in terms of the number of pentamers associated with a maximum of four specific links that can be accommodated. 

The assembly kinetics is characterized by a delay time $t_d$ for the onset of particle production and a thermodynamic relaxation time $t_r$. When the solution concentration $c_0$ of pentamers is lowered, assembled particles are stable against disassembly by thermal fluctuations on time scales shorter than $t_r$. If the energy scale of the binding energy is large compared to the thermal energy and if the ratio $|\epsilon_1|$ between specific and non-specific affinity of links of the genome molecules for the pentamer edges is small compared to one then there is a pronounced separation in time scale between particle assembly at high $c_0$ and particle disassembly at low $c_0$. We carried out packaging selection contests between molecules with different MLD and $N_P$. The selectivity is a purely kinetic effect, based on minor but systematic differences between the minimum energy assembly pathways of the different genome molecules.  For small values of $|\epsilon_1|$ molecules with small MLD and large $N_P$ outcompete molecules with large MLD and small $N_P$. For increasing $|\epsilon_1|$, genome selectivity increases as well but if $|\epsilon_1|$ becomes comparable to one then assembly produces a polydisperse solution of partially assembled particles that readily disintegrate when the pentamer solution is reduced. In addition, for large $|\epsilon_1|$ the assembly energy profiles become increasingly complex, leading to the failure of the nucleation-and-growth scenario. Another interesting consequence of increasing $|\epsilon_1|$ is a change from mechanically rigid to mechanically flaccid assembly intermediates.

In the Introduction we mentioned a number of experimental observations that motivated the construction of the model. The first was an in-vitro study of the co-assembly of CCMV with non-CCMV RNA molecules \cite{Comas-Garcia}. When the RNA-to-protein mixing ratio was low, virus-like particles (VLPs) formed with excess proteins, in agreement with simple arguments based on phase-coexistence. When the mixing ratio exceeeded a threshold, the virus-like particles were replaced by disordered RNA-protein aggregates instead of the expected coexistence of VLPs with excess RNA. Moreover, the threshold point separating the two regimes was well below the stoichiometric ratio. Can this be understood in the light of our results? When the mixing ratio (i.e., the depletion factor $D$) is increased at fixed total protein concentration $c_0$ in Fig. \ref{fig:PD} then for smaller$D$ practically all tree molecules are encapsidated (i.e., to the left of the blue line in Fig. \ref{fig:PD}) while larger values of $D$ an increasing fraction of the tree molecules are not encapsidated. The transition point separation the two regimes drops below the stoichiometric ratio when $c_0$ approaches the CAC. The authors of the experimental study \cite{Comas-Garcia} relate this displacement away from the stoichiometric ratio to electrostatic effects, which are not included in our model. The model shows that there is a separate mechanism that could produce the same effect. An experimental study in which the capsid protein concentration is reduced towards the CAC, perhaps accompanied by changes in the salinity, may be able to distinguish between the two mechanisms. Next, Fig. \ref{fig:D} shows that with increasing $D$, partially assembled capsids start to appear as observed experimentally. This would seem to be consistent with the proposed model but this polydispersity only becomes a dominant effect if $|\epsilon_1|$ approaches one (see Fig. \ref{fig:D} bottom). In that case assembled particles should be thermodynamically unstable, as discussed in section IV-C. In other words, they could not be true VLPs. This certainly would be an interesting result if it were confirmed experimentally.  A second interesting observation was the fact that for increasing values of the ratio between the RNA-protein and the protein-protein affinities, virus-like particles were replaced by disordered aggregates \cite{Garmann}. In our model, this ratio is represented by $|\epsilon_1|$. For increasing $|\epsilon_1|$ fully assembled shells indeed are replaced by partially assembled shells. Depending on the MLD and $N_P$ numbers, the assemblies can be mechanically unstable. A third observation concerned the fact that asymmetric reconstruction of the MS2 virus showed that a subsection of the RNA genome reproducibly associated with a compact cluster of capsid proteins \cite{Dykeman2011}. Our model reproduces this observation for genome molecules with small MLD and large $N_P$. The initial assembly is a compact cluster composed of maximally wrapped pentamers (see Fig.3, top left). 

While the model qualitatively reproduces these observations there is an important quantitative difference. As compared with the experiments of ref. \cite{Comas-Garcia} on CCMV, the model appears to underestimate the ability of increased RNA-to-protein mixing ratios to suppress the assembly of complete capsids. While this could be something specific for CCMV, we believe that this due to the fact that the model underestimates the entropy of a partial assembly. This is particularly obvious for flaccid intermediates that are expected to undergo strong thermal conformational fluctuations that are not represented in the model. While it may be possible to include conformational fluctuations in the model, a simpler route may be to carry out Brownian Dynamics simulations such as those of Ref. \cite{Perlmutter2014} but then for a system with pentamers that interact with tree molecules. Another natural extension of the model would be to larger viruses. The Zlotnick Model can be viewed as a representation of the smallest capsids, known as $T=1$ capsids, in which all capsid proteins have the same local packing organization. The CCMV and MS2 viruses that are an important testing ground for physical theories of viral assemblies are T=3 viruses that have a more complex architecture.

\begin{acknowledgements}
We would like to thank Alexander Grosberg for drawing our attention to spanning trees in the context of virus structure and assembly. We benefitted from discussions with Justin Little, Chen Lin, Zach Gvildys and William Vong. RB would like to thank the NSF-DMR for continued support under CMMT Grant No.1836404.
\end{acknowledgements}

\clearpage

\begin{appendix}
\section{Demonstration hat the smallest MLD for spanning trees on the dodecahedron is nine} \label{app:A}
\begin{figure}[htbp]
\begin{center}
\includegraphics[width=2.5in]{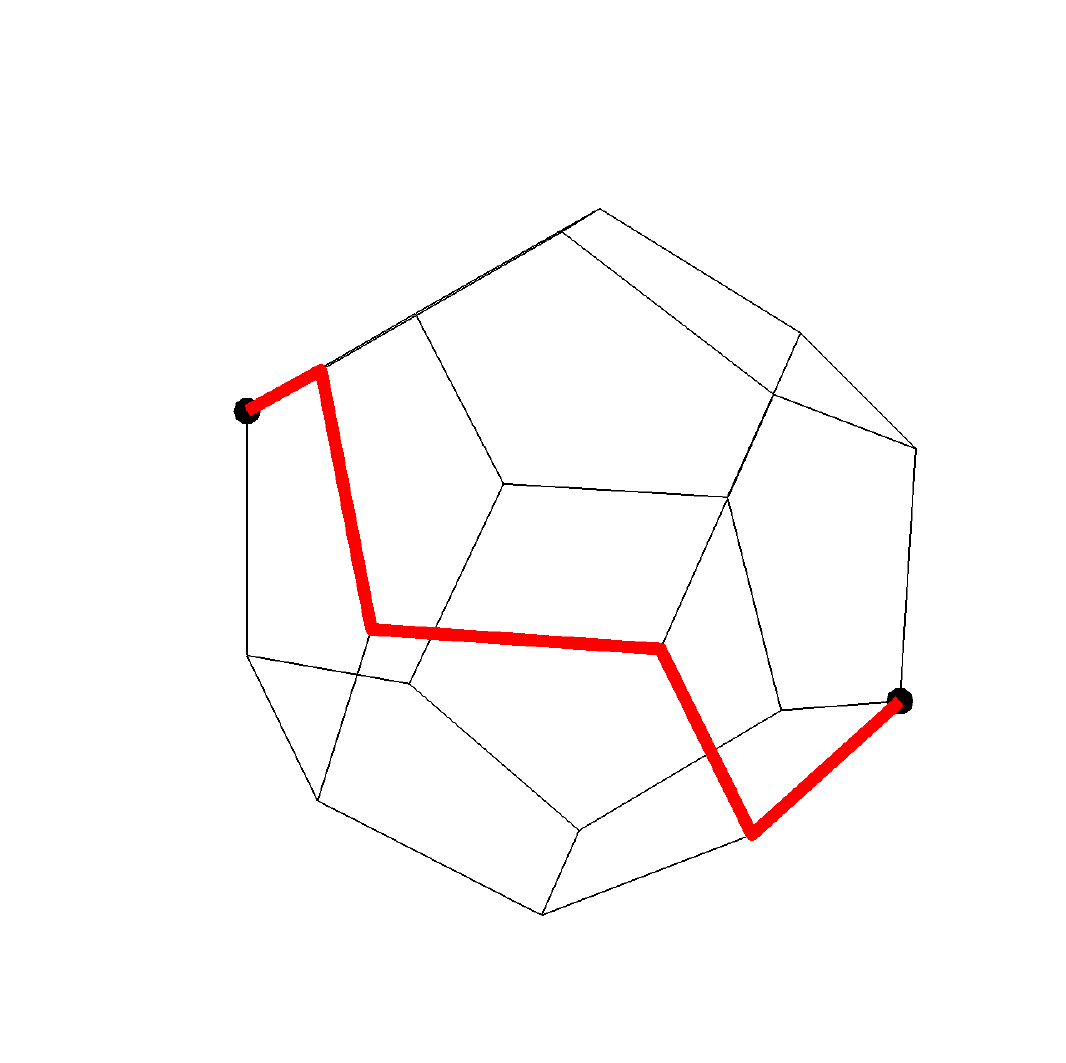}
\caption{Two maximally separated vertices on the dodecahedron and one of the 12 shortest paths consisting of five edges that join them.}
\label{fig:app1}
\end{center}
\end{figure}

We begin by noting that for every vertex on the dodecahedron there is a vertex on the opposite side of the polyhedron that is a ladder distance five away. That is, getting from one of the two vertices to the other requires traversing at least five edges. Figure \ref{fig:app1} shows such a path.
For each such pairs of vertices there are 12 minimal paths. 

\begin{figure}[htbp]
\begin{center}
\includegraphics[width=2.5in]{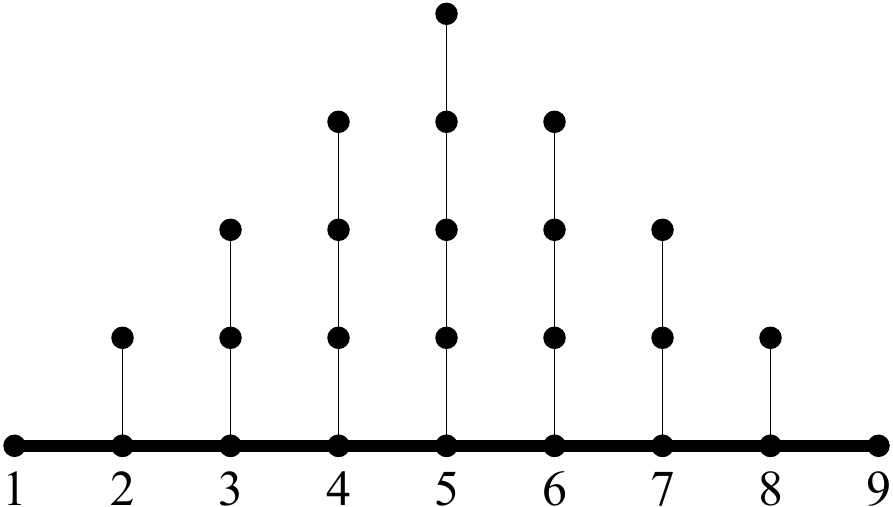}
\caption{A ladder distance 8 path in the hypothetical MLD 8 spanning tree on the dodecahedron. The path is shown as a thick line, and the nine vertices are labeled for easy reference. The thinner vertical lines represent longest allowed paths branching off the ladder distance 8 path. }
\label{fig:app2}
\end{center}
\end{figure}

Now, assume that there is a spanning tree with MLD 8. In such a case, we can pick out a path of ladder distance eight in that tree. All other elements of the tree will consist of trees that branch out from that path. Figure \ref{fig:app2} is a figurative depiction of the path along with the longest allowed branch sprouting off each vertex on that path. The likelihood of branching off those ``side branch'' paths is ignored; such branching does not alter the argument below. 

Consider  first the central vertex on the ladder distance eight path, labeled 5 in Fig. \ref{fig:app2}. The side path with ladder distance four is the longest that can attach to it. A longer path increases the MLD of the tree. Clearly, there is no possibility of reaching a point a ladder distance five from vertex 5 along any path with ladder distance four, so the path shown cannot connect the central vertex to the vertex a distance five away from it.  Next, consider the two sites flanking the central vertices, labeled 4 and 6. Attached to each is the longest possible path branching out from them, Such a path has ladder distance three. If either of these paths reached to the vertex a ladder distance five away from the central vertex, then there would be a ladder distance four (or less) path from that vertex through one of the flanking vertices to the maximally separated vertex, and we know that no such path exists. We can continue this argument to encompass all allowed paths sprouting from vertices on the chosen path. Thus, there is a vertex on the dodecahedron that cannot be a part of the MLD  8 tree containing this path. Consequently no tree with MLD 8 can be a spanning tree on the dodecahedron. The argument above can clearly be applied to the possibility of a spanning tree with MLD less than eight. That there is a spanning tree with MLD 9 is readily established by construction. 
\end{appendix}

\end{document}